\documentclass[runningheads]{llncs}
\usepackage[T1]{fontenc}
\usepackage{multirow}
\usepackage[title]{appendix}
\usepackage{subfig}
\usepackage{graphicx}
\usepackage{wrapfig}
\usepackage{blindtext}
\usepackage{amsmath}
\usepackage{amssymb}
\usepackage{booktabs}
\usepackage{siunitx}
\usepackage[misc,geometry]{ifsym}
\usepackage[hidelinks,colorlinks=true,allcolors=blue]{hyperref}
\usepackage{color}
\usepackage{algorithm,algpseudocode}

\newcommand{\etal}{\textit{et al.}}
\newcommand{\rvlcdip}{RVL-CDIP~\cite{harley2015icdar}}
\hypersetup{final}
\makeatletter
\renewcommand\section{\@startsection{section}{1}{\z@}%
	{-8\p@ \@plus -4\p@ \@minus -4\p@}%
	{6\p@ \@plus 4\p@ \@minus 4\p@}%
	{\normalfont\large\bfseries\boldmath
		\rightskip=\z@ \@plus 8em\pretolerance=10000 }}
\renewcommand\subsection{\@startsection{subsection}{2}{\z@}%
	{-8\p@ \@plus -4\p@ \@minus -4\p@}%
	{6\p@ \@plus 4\p@ \@minus 4\p@}%
	{\normalfont\normalsize\bfseries\boldmath
		\rightskip=\z@ \@plus 8em\pretolerance=10000 }}
\renewcommand\subsubsection{\@startsection{subsubsection}{3}{\z@}%
	{-6\p@ \@plus -4\p@ \@minus -4\p@}%
	{-0.5em \@plus -0.22em \@minus -0.1em}%
	{\normalfont\normalsize\bfseries\boldmath}}

\makeatother
\begin{document}
\title{
	DP-DocLDM: Differentially Private Document Image Generation using Latent Diffusion Models
}
\titlerunning{DP-DocLDM: Differentially Private Document Image Generation}
\author{
	Saifullah Saifullah(\Letter)\inst{1,2} \orcidID{0000-0003-3098-2458} \and
	Stefan Agne\inst{1,3} \orcidID{0000-0002-9697-4285} \and
	Andreas Dengel\inst{1,2} \orcidID{0000-0002-6100-8255} \and
	Sheraz Ahmed\inst{1,3} \orcidID{0000-0002-4239-6520}}
\authorrunning{S. Saifullah \textit{et al.}}
\institute{
	Deutsches Forschungszentrum für Künstliche Intelligenz GmbH (DFKI), Trippstadter Straße 122,
	67663 Kaiserslautern, Germany\\\email{\{firstname.lastname\}@dfki.de}\\ \and
	Department of Computer Science, RPTU Kaiserslautern-Landau, Erwin-Schrödinger-Straße 52, 67663 Kaiserslautern, Germany\and
	DeepReader GmbH, 67663 Kaiserlautern, Germany\\
}
\maketitle
\begin{abstract}
As deep learning-based, data-driven information extraction systems become increasingly integrated into modern document processing workflows, one primary concern is the risk of malicious leakage of sensitive private data from these systems. While some recent works have explored Differential Privacy (DP) to mitigate these privacy risks, DP-based training is known to cause significant performance degradation and impose several limitations on standard training procedures, making its direct application to downstream tasks both difficult and costly.
In this work, we aim to address the above challenges within the context of document image classification by substituting real private data with a synthetic counterpart. In particular, we propose to use conditional latent diffusion models (LDMs) in combination with differential privacy (DP) to generate class-specific synthetic document images under strict privacy constraints, which can then be utilized to train a downstream classifier following standard training procedures.
We investigate our approach under various pretraining setups, including unconditional, class-conditional, and layout-conditional pretraining, in combination with multiple private training strategies such as class-conditional and per-label private fine-tuning with DPDM and DP-Promise algorithms.
Additionally, we evaluate it on two well-known document benchmark datasets, RVL-CDIP and Tobacco3482, and show that it can generate useful and realistic document samples across various document types and privacy levels ($\varepsilon \in \{1, 5, 10\}$).
Lastly, we show that our approach achieves substantial performance improvements in downstream evaluations on small-scale datasets, compared to the direct application of DP-Adam.
To the best of the authors' knowledge, this is the first work to explore differentially private document generation via latent diffusion for document analysis. The source code for this work has been made publicly available at: \url{https://github.com/saifullah3396/dpdocldm.git}
\end{abstract}
\section{Introduction}
Modern breakthroughs in deep learning (DL)~\cite{resnet,bert-devlin2019,dosovitskiy2021an} have revolutionized the automated processing of business documents, leading to significant progress across a variety of document analysis tasks, such as document image classification~\cite{Afzal2017-deep-cnns,Saifullah2022-docxclassifier,doc-vit,layoutlmv3-Huang2022}, key information extraction~\cite{tilt-Powalski2021,layoutlmv3-Huang2022,formnet-Lee2022}, and document layout analysis~\cite{layout-shen2021,layoutlmv3-Huang2022}.
However, given the recent alarms raised about potential data breaches~\cite{memorization-Carlini2019,model-inversion-Coavoux2020,model-inversion-att-Fredrikson2015} and privacy risks~\cite{attacks-survey-Al-Rubaie2019,priv-survey-Hu2023,membership-inf-att-Shokri2017} surrounding DL-powered systems, deploying such models on real-world business documents---which typically contain a wealth of sensitive and confidential information---carries significant legal and ethical risks, such as violating the GDPR~\cite{GDPR2016a} and the AI Act 2022.

Recent studies~\cite{memorization-Carlini2019,model-inversion-Coavoux2020,model-inversion-att-Fredrikson2015,image_from_grad-yin,text_from_llm-carlini} have shown that modern deep learning models provide little privacy for their training data, enabling malicious adversaries to extract various types of information from these models, such as the reconstruction of training images through gradient inversion~\cite{image_from_grad-yin}, the recovery of training data statistics~\cite{model-inversion-att-Fredrikson2015,model-inversion-Coavoux2020}, or the extraction of sample membership information~\cite{membership-inf-att-Shokri2017}.
To address these issues, Differential Privacy (DP)~\cite{dp-Dwork2006,dp-Dwork2014,dpsgd-Abadi2016} has emerged as the most prominent privacy framework, which, by definition, provides rigorous privacy guarantees for the training data and has been applied to a wide range of application domains~\cite{fl-mercier2021evaluating,dp-medical,dp-Wunderlich2022,dp-Li2021}.
However, training deep learning models with DP, especially in resource-constrained environments, remains a significant challenge due to its high computational costs~\cite{dp-Li2021,dp-papernot2020making}, excessive hyperparameter-tuning requirements~\cite{Saifullah2024,dp-Li2021}, and the substantial degradation of model utility~\cite{dp-papernot2020making,priv-documents-Basu2021,fl-mercier2021evaluating,dp-kie,dp-basu2022benchmarking} observed under private training. These issues have also been particularly highlighted in recent applications of DP to document analysis tasks, such as document classification~\cite{Saifullah2024}, document key information extraction~\cite{dp-kie} and document visual question answering~\cite{dp-docvqa}.

To circumvent these challenges, recent studies~\cite{harder2023pretrained,liew2022pearl,dockhorn2023differentially} have explored the potential of private generative approaches, such as differentially private generative adversarial networks (DP-GANs)~\cite{harder2023pretrained,liew2022pearl} or differentially private diffusion models (DPDMs)~\cite{dockhorn2023differentially}, to replace real sensitive data with private synthetic counterparts that follow the same distribution as the real data.
This not only enables the secure generation of additional data samples in domains with limited data availability but also makes it possible to train deep learning models for downstream tasks using standard training procedures.
However, despite significant progress in the field of document image analysis~\cite{Afzal2017-deep-cnns,Saifullah2022-docxclassifier,doc-vit,layoutlmv3-Huang2022,layout-shen2021,layoutlmv3-Huang2022}, and considerable recent attention given to non-private synthetic document data generation~\cite{ldms-doc,guan2024idnetnoveldatasetidentity,hamdani-doc,diffusion-layout-doc}, research focused on data privacy---particularly focused on privacy through synthetic counterparts---remains relatively scarce.

To address this research gap, we investigate private synthetic data generation for the task of document image classification, a vital component of automated document processing workflows. In particular, we analyze the potential of using latent diffusion models (LDMs)~\cite{rombach2021highresolution} in combination with differential privacy (DP)~\cite{dp-Dwork2006,dp-Dwork2014} to generate class-specific private synthetic document images that can substitute real private data for downstream tasks, in this case, document image classification.
To this end, we first pretrain a class- and layout-conditioned latent diffusion model on a large public document image dataset, and then transfer this knowledge via differentially private fine-tuning to generate documents for the smaller private datasets. Finally, we replace the real private data with synthetic data for downstream evaluation.

Overall, our main contributions are as follows:
\begin{itemize}
	\item We propose DP-DocLDM, a novel approach that combines conditional latent diffusion models (LDMs)~\cite{rombach2021highresolution} with differential privacy (DP)~\cite{dp-Dwork2006,dp-Dwork2014} to generate private synthetic document images. To the best of the authors' knowledge, this is the first work to explore differentially private synthetic document image generation with diffusion models.
    \item We present an extensive benchmarking of the proposed strategy across various pretraining setups, including unconditional, class-conditional, and layout-conditional pretraining, in combination with various private training strategies, such as class-conditional and per-label private fine-tuning with DPDM~\cite{dockhorn2023differentially} and  DP-Promise~\cite{dppromise}.
	\item
    Lastly, through a rigorous qualitative and quantitative assessment of our approach on two well-known document benchmark datasets, \rvlcdip{} and Tobacco3482, we demonstrate its ability to generate useful and realistic document samples across various document types and privacy levels ($\varepsilon\in\{1,5,10\}$).
    Furthermore, we show that our approach achieves significant performance improvements in downstream evaluations on small-scale datasets, compared to the direct application of DP-SGD~\cite{Saifullah2024}.
\end{itemize}
\section{Related Work}
\subsection{Differential Privacy in Document AI}
Differential Privacy (DP)~\cite{dp-Dwork2006,dp-Dwork2014} and its various adaptations~\cite{dpsgd-Abadi2016,local-dp-cape-plant2021,local-mdp-Feyisetan2019} have gained significant attention in recent years for ensuring privacy within the field of Document AI, encompassing both textual and visual contexts.

In the textual context, Li~\etal~\cite{dp-Li2021} proposed optimal training hyperparameters for the private fine-tuning of large language models in text classification and generation tasks. Hoory~\etal~\cite{dp-Hoory2021} investigated global DP for named-entity recognition in medical documents and introduced a domain-specific private vocabulary for training BERT-based models. Basu~\etal~\cite{dp-basu2022benchmarking} investigated global DP for the classification of financial documents. McMahan~\etal~\cite{fl-rnn-McMahan2017LearningDP} proposed large-scale distributed training of recurrent neural networks (RNNs) for textual data under global client-level differential privacy. For a comprehensive overview of the applications of local and global DP in textual documents, we refer the reader to a related survey~\cite{priv-survey-Hu2023}.

In visual contexts, privacy has been relatively underexplored and has only recently gained attention. Saifullah~\etal~\cite{Saifullah2024} presented a comprehensive study benchmarking several state-of-the-art privacy-preserving methods, including differential privacy, federated learning, and homomorphic encryption, for document image classification.
In another recent study, Saifullah~\etal~\cite{dp-kie} explored the combination of differential privacy and federated learning in a multi-modal setting for the task of document key information extraction.
Similarly, Tito~\etal~\cite{dp-docvqa} proposed a large-scale differential privacy training setup for document visual question answering (VQA) tasks.
Most recently, Guan~\etal~\cite{guan2024idnetnoveldatasetidentity} investigated the applications of traditional masking approaches and the PixelDP~\cite{pixeldp} algorithm on sensitive identity documents.

\subsection{Privacy through Generative Modeling}
Privacy through generative modeling has recently gained considerable traction due to the superior generative capabilities of diffusion models~\cite{dockhorn2023differentially,tsai2024differentiallyprivatefinetuningdiffusion,ghalebikesabi2023differentiallyprivatediffusionmodels}. While earlier works in this area relied predominantly on generative adversarial networks (GANs)~\cite{torkzadehmahani2019dp,NEURIPS2020_9547ad6b}, their well-known training instability issues, coupled with model utility challenges associated with DP, presented significant obstacles. In contrast, Differentially Private Diffusion Models (DPDMs)~\cite{dockhorn2023differentially,dppromise} have emerged as a promising alternative, offering both stable training and improved sample quality under private training.
A number of recent works have also extended DPDMs~\cite{dockhorn2023differentially} to latent-space diffusion, achieving promising results in high-resolution image synthesis under rigorous privacy constraints~\cite{tsai2024differentiallyprivatefinetuningdiffusion,ghalebikesabi2023differentiallyprivatediffusionmodels}.

\subsection{Diffusion Models for Synthetic Document Generation}
Diffusion models (DMs)~\cite{ho2020denoisingdiffusionprobabilisticmodels} have also recently gained significant popularity for non-private document image synthesis. For instance, Tanveer~\etal~\cite{ldms-doc} recently proposed text-based layout conditioning for generating pre-annotated synthetic document images for document layout analysis. Similarly, Hamdani~\etal~\cite{hamdani-doc} proposed mask-based layout-conditioning for document table generation. Guan~\etal~\cite{guan2024idnetnoveldatasetidentity} proposed to incorporate latent diffusion models~\cite{rombach2021highresolution} for generating fake identity images in synthetic identity documents. In a slightly different direction, Liu~\etal~\cite{diffusion-layout-doc} proposed diffusion-based layout generation for documents.

\section{Preliminaries}
\subsection{Diffusion Models}
Diffusion models~\cite{sohldickstein2015deepunsupervisedlearningusing,ho2020denoisingdiffusionprobabilisticmodels} have gained significant traction in recent years due to their excellent generative capabilities~\cite{ho2022classifierfreediffusionguidance,dhariwal2021diffusionmodelsbeatgans},
the core working principle behind which is to first gradually transform the real data distribution into a prior distribution (typically Gaussian) through a forward diffusion process, and then gradually undo this transformation to recover the original data distribution.
In particular, for any data point sampled from the real distribution $x_0\sim q(x_0)$, the forward diffusion process $q(x_1,\dots,x_T|x_0)$ defines a fixed Markov chain $\{x_1,\dots,x_T\}$ by iteratively perturbing it with Gaussian noise $\epsilon \sim \mathcal{N}(0, I)$:
\begin{align}
    q(x_t|x_{t-1}) &= \mathcal{N}(x_t; \sqrt{\beta_t}x_{t-1}, (1-\beta_t)\mathbf{I})
\end{align}
where $\beta_{1,\dots,T}$ is a fixed variance schedule that controls the magnitude of noise added at each timestep and is defined such that $q(x_T)\sim\mathcal{N}(0, I)$. A notable property of the forward process is that it enables sampling $x_t$ at any arbitrary timestep $t$ in closed form:
\begin{align}
    q(x_t|x_0) &= \mathcal{N}(x_t; \sqrt{\bar{\alpha}_t}x_{t-1}, (1-\bar{\alpha}_t)\mathbf{I})\\
    x_t &= \sqrt{\bar{\alpha}_t}x_{0} + \sqrt{1-\bar{\alpha}_t}\epsilon,\quad \epsilon\sim\mathcal{N}(0, I)
    \label{eq:forward-step}
\end{align}
where $\alpha_t:=1 - \beta_t$ and $\bar{\alpha}_t:=\prod_{k=1}^t\alpha_k$.
In the reverse diffusion process, a denoising diffusion model, $p_\theta$ with parameters $\theta$ learns to approximate each backward transition step $q(x_{t-1}|x_{t})$ as $p_\theta(x_{t-1}|x_t)$ by predicting the approximate mean $\mu_\theta(x_t, t)$ and covariance $\sigma_\theta(x_t, t)$ of the transition, which is then repeatedly applied to generate new samples $x_0\sim p_\theta(x_0)$:
\begin{align}
    p_\theta(x_{t-1}|x_t) &:=\mathcal{N}(x_{t-1}; \mu_\theta(x_t, t), \Sigma_\theta(x_t, t))
    \label{eq:reverse-step}
\end{align}
In practice, instead of predicting the mean $\mu_\theta(x_t, t)$ directly, the diffusion model is typically trained to predict the source noise $\epsilon_\theta(x_t, t)$ added at each timestep with the following simplified loss objective:
\begin{equation}
    \label{eq:loss}
    \mathbb{E}_{t \sim [1,T],\epsilon\sim\mathcal{N}(0, I),x_{0}\sim q(x_0)}||\epsilon - \epsilon_\theta(x_t, t)||^2
\end{equation}
whereas the mean $\mu_\theta(x_t, t)$ is indirectly obtained as:
\begin{equation}
    \mu_\theta(x_t, t) = \frac{1}{\sqrt{\alpha_t}}(x_t - \frac{\beta_t}{\sqrt{1-\bar{\alpha}_t}}\epsilon_\theta(x_t,t)).
\end{equation}

\subsection{Differential Privacy (DP)}
\label{sec:dp}
Differential Privacy (DP) is a well-known privacy framework that formalizes the information release from any randomized algorithm and offers strong theoretical guarantees for data privacy.
In this work, we focus on the sample-level privacy of the training dataset under the global $(\varepsilon,\delta)$-DP, formally defined as follows:
\begin{definition}
	A randomized algorithm $\mathcal{M} : \mathcal{D} \rightarrow \mathcal{R}$ with domain $\mathcal{D}$ and range $\mathcal{R}$ is $(\varepsilon,\delta)$-differentially private if for all $S \subseteq \mathcal{R}$ and for all datasets $D,D' \in \mathcal{D}$ that differ at most in one record:
	\begin{equation*}
		\mathbb{P}(\mathcal{M}(D)\in S)\leq e^{\varepsilon}\mathbb{P}(\mathcal{M}(D')\in S)+\delta
	\end{equation*}
\end{definition}
Intuitively, applying Differential Privacy (DP) to an algorithm $\mathcal{M}$ ensures that its output distribution on two datasets that only differ in a single sample does not vary significantly. Whereas, the degree of this variability is controlled by the privacy parameters $(\varepsilon, \delta)$, with lower values of these parameters indicating stronger privacy constraints.

\subsubsection{DP-SGD/Adam} For applying DP to deep neural networks, differentially private stochastic gradient descent (DP-SGD)~\cite{dpsgd-Abadi2016} is the primary algorithm. The core idea behind DP-SGD~\cite{dpsgd-Abadi2016} is to add noise to the per-sample gradients of the model during model optimization in order to minimize its dependence on individual samples. In particular, to satisfy the constraints of global $(\varepsilon$, $\delta)$-DP, the DP-SGD~\cite{dpsgd-Abadi2016} algorithm first clips the per-sample gradients to a fixed bound $C$, and adds Gaussian noise $n\sim \mathcal{N}(0, \sigma^2C^2)$ to the sum of the gradients before each gradient optimization step. Here, $\sigma$ is the noise multiplier that determines the overall privacy strength with higher values resulting in stronger privacy constraints (lower $\varepsilon$). In practice, $\sigma$ is chosen such that over the complete training cycle, a required privacy budget $(\varepsilon$, $\delta)$ is spent. This is done by tracking the privacy loss using privacy accountants~\cite{dp-rdp-mironov,dp-gdp-koskela2022individual}.

\subsubsection{DPDM} DPDM~\cite{dockhorn2023differentially} is an extension of DP-SGD~\cite{dpsgd-Abadi2016} that incorporates noise multiplicity $\eta$ in private optimization. Essentially, noise multiplicity involves sampling multiple diffusion timesteps for each sample in a batch, repeated $\eta$ times, and then averaging the per-sample gradients over these timesteps before applying DP-SGD~\cite{dpsgd-Abadi2016} update. This allows for significant improvements in model performance without any additional privacy loss. The full pseudocode of the DPDM~\cite{dockhorn2023differentially} algorithm is provided in Appendix~\ref{app:privacy-algorithms}.

\subsubsection{DP-Promise} DP-Promise~\cite{dppromise} is another algorithm recently proposed for training diffusion models under differential privacy (DP).
Since diffusion models themselves also add noise to the input during training, which evolves into pure noise $\epsilon \sim \mathcal{N}(0, I)$ as the timestep $t$ $\rightarrow$ $T$, the learned transitions $p_\theta(x_{t-1}|x_t)$ in the high-noise regimes lose most of the information about the original sample.
Wang~\etal~\cite{dppromise} showed that one can use this forward noise addition process to also directly ensure differential privacy (DP) for the sample training. However, this is only possible in high-noise regimes ($t \rightarrow T$). Overall, DP-Promise~\cite{dppromise} works by first training the diffusion model for a few epochs only on timesteps $t$ close to $T$ and then privately training the model using the standard DP-SGD algorithm~\cite{dpsgd-Abadi2016}.
In this work, we investigate both DP-SGD~\cite{dpsgd-Abadi2016} and DP-Promise~\cite{dppromise} for private training of latent diffusion models.

\section{DP-DocLDM: Differentially Private Document Image Generation using Latent Diffusion Models}
\begin{figure}[t!]
    \centering
    \includegraphics[width=0.8\textwidth]{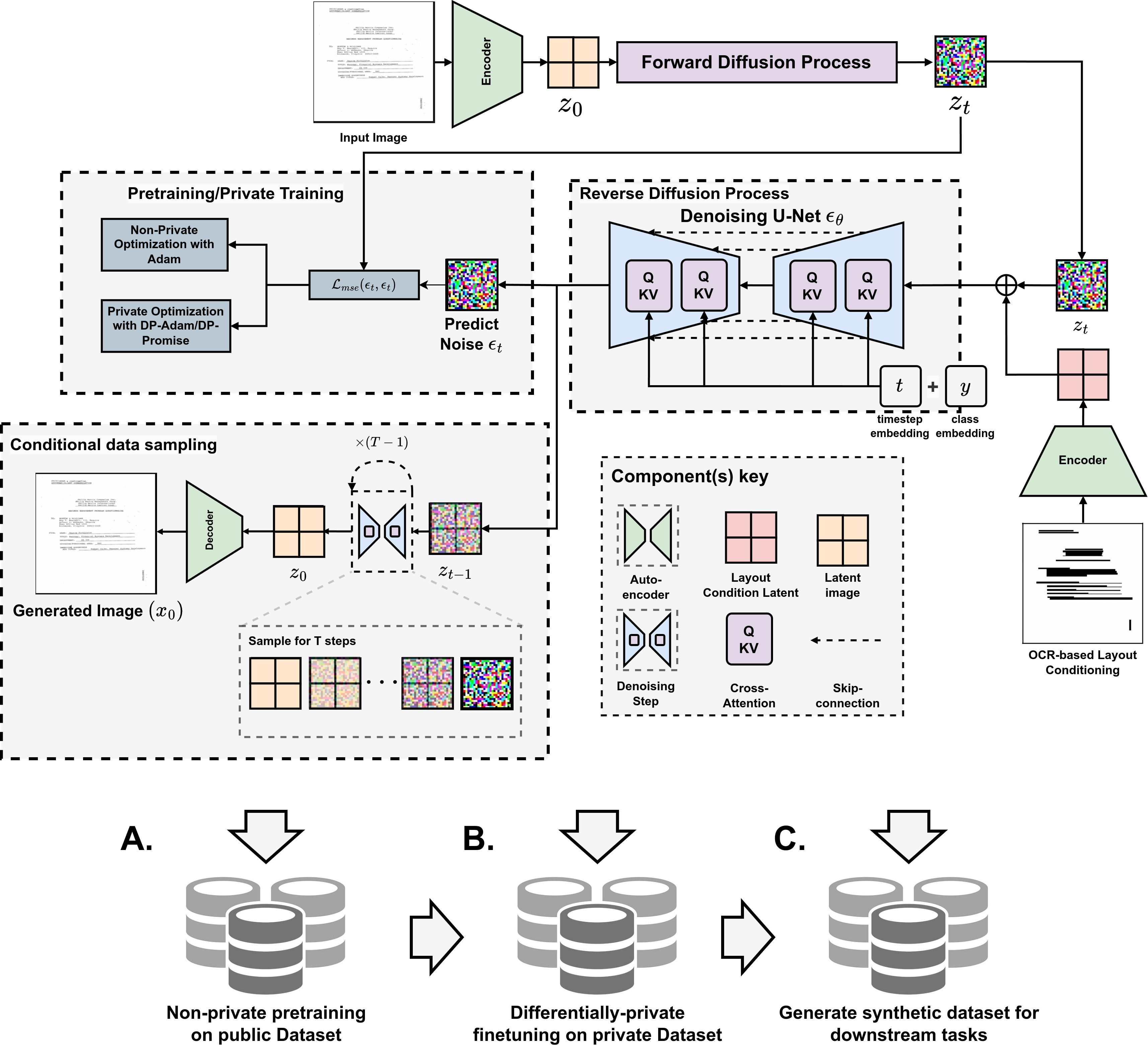}
    \caption{
        Overview of the proposed approach, DP-DocLDM. The input image is first converted into a noisy latent representation using forward diffusion.
        Simultaneously, an OCR-based binary layout mask is encoded into a separate latent image and concatenated with the noisy image. The model is then trained to predict the noise added to the sample. We first pretrain the model on $\mathcal{D}_\text{public}$ (A), followed by private fine-tuning on $\mathcal{D}_\text{private}$ (B). Finally, the privately trained model is used to generate a synthetic counterpart for $\mathcal{D}_\text{private}$ (C).
    } \label{fig:approach}
    \vspace{-1em}
\end{figure}
In this section, we present the details of our proposed approach, DP-DocLDM, for replacing private document datasets with synthetically generated counterparts using differentially private latent diffusion models.
As outlined in Fig.~\ref{fig:approach}, our method involves generating private document images in three steps: (A) non-private pretraining of a class- and layout-conditioned latent diffusion model on a large public document image dataset $\mathcal{D}_{\text{public}}$, (B) private fine-tuning of the model on the private document dataset $\mathcal{D}_{\text{private}}$, and (C) sampling from the private model to generate a new synthetic dataset, which is then substituted for the private dataset in downstream evaluation.
\subsection{Non-private Pretraining on Public Dataset}
\label{sec:pretraining}
Training a diffusion model with DP-SGD~\cite{dpsgd-Abadi2016} on high-resolution images (sizes $>256\times 256$) is computationally expensive. Therefore, in this work, we instead train a private latent diffusion model (LDM)~\cite{rombach2021highresolution} as done in previous works~\cite{ghalebikesabi2023differentiallyprivatediffusionmodels,liu2024differentially}.
Latent diffusion models (LDMs)~\cite{rombach2021highresolution} learn to generate new samples in the compressed latent space of a pretrained autoencoder $z=\mathcal{D}(\mathcal{E}(x))$, which down-scales the high-resolution input images $x\sim \mathbb{R}^{C\times H \times W}$ into a compressed latent representation $z\sim \mathbb{R}^{C_z\times \frac{H}{f} \times \frac{W}{f}}$, where $f$ is the down-scaling factor.
Using the latent-space representation of the data not only reduces training and data generation costs but also facilitates the model to focus only on high-level semantic features instead of high-frequency details in the pixel space.
Following Romback~\etal~\cite{rombach2021highresolution}, we pretrain a KL-reg autoencoder variant on the public document image dataset $\mathcal{D}_{\text{public}}$ using a combination of adversarial and perceptual losses. It is worth noting that since the autoencoder is trained on the public dataset, it does not introduce any privacy risks.
Subsequently, we pretrain the latent diffusion model on $\mathcal{D}_{\text{public}}$ by minimizing the loss objective in Eq.~\ref{eq:loss} in the latent space:
\begin{equation}
    \min_\theta \mathbb{E}_{t \sim [1,T],\epsilon\sim\mathcal{N}(0, I),z_{0}\sim \mathcal{E}(\mathcal{D}_{\text{public}})}||\epsilon - \epsilon_\theta(z_t, t)||^2
\end{equation}
Note that the above loss objective is for unconditional case.
Apart from exploring the unconditional pretraining on original public dataset ({Uncond.+Unaug.}), we explore 3 additional options: (1) pretraining on augmented dataset ({Uncond.+Aug.}), (2) pretraining with class-conditioning ({Class Cond.}), and (3) pretraining with a combination of class and layout conditioning ({Layout+Class Cond.}).
We explore pretraining on the augmented dataset mainly to assess how a wide discrepancy between the public data ($\mathcal{D}_{\text{public}}$) and the private dataset $\mathcal{D}_{\text{private}}$ may affect model performance.
We explore class-conditioning~\cite{ho2020denoisingdiffusionprobabilisticmodels,ho2022classifierfreediffusionguidance} to assess whether prior class information can help guide the private fine-tuning process.
Finally, since class information is generally not available in public pretraining datasets, we propose using OCR-based layout-conditioning to provide additional information about the document during the generation process.
To this end, we first extract the text bounding-boxes from the data using the publicly available Tesseract OCR library~\cite{TessOverview}, and generate a binary mask image (as shown in Fig.~\ref{fig:approach}) that holds the layout information of the document.
In this work, we use line-level segmentation to generate the layout masks. With these conditioning mechanisms, the loss objective in Eq.~\ref{eq:loss} can be updated as follows:
\begin{equation}
    \min_\theta \mathbb{E}_{t \sim [1,T],\epsilon\sim\mathcal{N}(0, I),z_{0}\sim \mathcal{E}(\mathcal{D}_{\text{public}})}||\epsilon - \epsilon_\theta(z_t, t, c, \mathcal{E}(m))||^2
\end{equation}
where $c$ denotes the class embedding, which is added to the timestep embedding $t$, and $\mathcal{E}(m)$ denotes the latent-space encoding of the input layout mask $m$, which is concatenated with the noisy latent image $z_t$. Note that we use the same encoder, $\mathcal{E}$, for both the layout mask $m$ and the input image $x$.
\subsection{Differentially Private Fine-tuning on Private Dataset}
\label{sec:private-training}
To generate a synthetic counterpart of the private dataset $\mathcal{D}_{private}$  with privacy constraints, we privately fine-tune the pretrained diffusion model directly on $\mathcal{D}_{private}$ using DPDM~\cite{dockhorn2023differentially} and DP-Promise~\cite{dppromise} (with Adam optimizer).
In particular, for all 4 pretraining setups outlined in Section~\ref{sec:pretraining}, we fine-tune the model on the private dataset under 2 different setups. (1) In the first setup, we privately fine-tune the models with class conditioning~\cite{ho2020denoisingdiffusionprobabilisticmodels,rombach2021highresolution} and classifier-free guidance~\cite{ho2022classifierfreediffusionguidance} similar to previous works~\cite{ghalebikesabi2023differentiallyprivatediffusionmodels,liu2024differentially}.
(2) In the second setup, we drop the class conditioning entirely and instead fine-tune an independent private LDM for each document class subset.
There are several reasons for adopting this approach. First, document datasets are well-known for exhibiting extremely high intra-class variance and inter-class similarity~\cite{Afzal2017-deep-cnns,Saifullah2022-docxclassifier}.
This means that two samples from entirely different document classes may be perceptually similar, while two samples from the same document class may look completely different.
This makes it challenging for a diffusion model to correctly learn the class distributions under private training.
Therefore, by using unconditional training for each document class, we allow the model to focus solely on learning the semantic features of the given class, rather than simultaneously learning to discriminate between classes.
Note that for both private fine-tuning setups, in cases where layout conditioning is used during pretraining, we also apply layout conditioning during private training by extracting the layouts from the private dataset.

Finally, once the models have been fine-tuned in a private manner, we generate new samples for each target document class by conditioning on the target classes in the first setup and by using the individual class-specific LDMs in the second setup. Additionally, for both setups, when layout conditioning is required, we apply layout conditioning during sampling using layouts extracted from the training set of the private dataset.

\section{Experiments and Results}
\subsection{Datasets}
\subsubsection{Public dataset}
For public pretraining of both the autoencoder and the diffusion model, we utilize the large-scale public document dataset IIT-CDIP Test Collection 1.0~\cite{iitcdip}.
The IIT-CDIP Test Collection 1.0~\cite{iitcdip} contains a total of 11 million document images, approximately 4.5 million of which are weakly labeled with multiple class labels per document.
For pretraining the autoencoder, we use the entire set of 11 million document images. However, for both conditional and unconditional pretraining of the diffusion model, we filter the dataset to the 80 most frequent classes that contain at least 1,000 samples each for training.
Moreover, we observed that the class distribution within the dataset is highly skewed, with millions of documents belonging to some classes and only a few thousand to others. To address this imbalance, we apply weighted random sampling in all pretraining experiments.

\subsubsection{Private datasets}
For private training, we evaluate our approach on two widely used document benchmark datasets: \rvlcdip{} and Tobacco3482\footnote{https://www.kaggle.com/datasets/patrickaudriaz/tobacco3482jpg}, both of which are frequently employed for document classification tasks~\cite{layoutlmv3-Huang2022,harley2015icdar,Saifullah2022-docxclassifier,ferrando2020-doc-class-4}.
\rvlcdip{} consists of 400,000 document images labeled into 16 different document categories.
It has a balanced class distribution and is split into training, validation, and testing sets with sizes of 320,000, 40,000, and 40,000 samples, respectively.
In contrast, Tobacco3482 is a relatively small-scale dataset with an imbalanced class distribution and 10 document categories. Following previous work~\cite{Saifullah2024}, we split this dataset using an 80/20 ratio into training, validation, and testing sets, with sizes of 2,504, 279, and 700 samples, respectively.

\subsection{Implementation details}
For all our experiments, we utilize two pretrained autoencoders, KL-F4 and KL-F8, each with an image downscaling factor of $f = 4$ and $f = 8$, respectively.
To predict the noise at each timestep, we adopt a UNet-based architecture proposed by Ho~\etal~\cite{ho2022classifierfreediffusionguidance}, which combines standard residual blocks~\cite{resnet} and attention blocks~\cite{bert-devlin2019} with a timestep embedding to generate its output.
For the forward diffusion process, we use standard linear noise scheduler with $(\beta_1,\beta_T)=(10^{-4},0.02)$ and $T=1000$.
To track the privacy loss $(\varepsilon, \delta)$ in private training setups, we use the R'enyi Differential Privacy (RDP) accountant for DPDM~\cite{dockhorn2023differentially} and the Gaussian Differential Privacy (GDP) accountant for DP-Promise~\cite{dppromise}, as originally proposed.
The noise multiplier $\sigma$ is computed based on the required privacy budget ($\varepsilon \in \{1, 5, 10\}$, $\delta = \frac{1}{|\mathcal{D}_{private}|}$) over a fixed number of training epochs (see Appendix~\ref{app:privacy-accounting} for more details).
Note that for per-label private training setups, the noise multiplier $\sigma$ is computed separately for each labeled subset.
All private training experiments are conducted with a gradient clipping threshold of $C = 0.01$, a batch size of $1096$, and a learning rate of $3 \times 10^{-4}$.
For private training on the \rvlcdip{} dataset, we train the models for 50 epochs with a noise multiplicity~\cite{dockhorn2023differentially} of $\eta \in \{1, 4\}$, while for the Tobacco3482 dataset, we train the models for 250 epochs with a noise multiplicity~\cite{dockhorn2023differentially} of $\eta = 32$.
For all private fine-tuning setups, we freeze the timestep embedding and train the rest of the model parameters ($\sim$32M). Moreover, in order to utilize classifier-free guidance~\cite{ho2022classifierfreediffusionguidance} in the class-conditional private setup, we replace the class labels with the null label with a probability of $0.1$ during training.
For generating data samples after private training, we use the standard DDPM sampler~\cite{ho2020denoisingdiffusionprobabilisticmodels} with a total of 200 inference steps.
\subsection{Evaluation protocol}
We conduct the evaluation of our approach in two stages: (1) comparative analysis of training strategies and (2) comprehensive final evaluation.
In the first stage, we evaluate our approach across the different pretraining setups described in Section~\ref{sec:pretraining}: (1) Uncond.+Unaug., (2) Uncond.+Aug., (3) Class Cond., and (4) Layout+Class Cond., in combination with the private training setups outlined in Section~\ref{sec:private-training}: (1) private fine-tuning with classifier-free guidance (CFG)~\cite{ho2022classifierfreediffusionguidance}, and (2) private fine-tuning on per-label subsets of the dataset, using both DPDM~\cite{dockhorn2023differentially} and DP-Promise~\cite{dppromise}.
In this stage, we train the diffusion models on \rvlcdip{} dataset with a target privacy budget of $\varepsilon=10$ across all settings and conduct downstream evaluation by training the ConvNeXt-B~\cite{liu2022convnet} model on 50K generated samples.

In the second stage, we use the best-performing approach from the first stage to conduct an evaluation under stricter privacy budgets of $\varepsilon \in \{1, 5, 10\}$ and perform downstream evaluation using three different classifiers: ResNet-50~\cite{resnet}, ConvNeXt-B~\cite{liu2022convnet}, and DiT-B~\cite{doc-vit}, where the DiT-B~\cite{doc-vit} model is pretrained in a self-supervised manner on the IIT-CDIP Test Collection 1.0~\cite{iitcdip} dataset. Additionally, we use a noise multiplicity~\cite{dockhorn2023differentially} of $\eta=4$ for \rvlcdip{} dataset in this stage.
For all downstream evaluations, we train the models using CutMix~\cite{yun2019cutmixregularizationstrategytrain}, Mixup~\cite{zhang2018mixupempiricalriskminimization}, a dropout rate of 0.5, and use soft-cross-entropy loss for model optimization.

\subsection{Stage 1: Comparative Analysis of Training Strategies}
\begin{table}[!t]
    \centering
    \resizebox{\textwidth}{!}{
        \begin{tabular}{llccccccccccccccc}
            \toprule
            &  \multicolumn{4}{c}{Uncond.+Unaug.} & \multicolumn{4}{c}{Uncond.+Aug.} & \multicolumn{4}{c}{Class Cond.} & \multicolumn{4}{c}{Layout+Class Cond.}\\ \cmidrule(lr){2-5} \cmidrule(lr){6-9} \cmidrule(lr){10-13}\cmidrule(lr){14-17}
            &  \multicolumn{2}{c}{KL-F8} & \multicolumn{2}{c}{KL-F4} & \multicolumn{2}{c}{KL-F8} & \multicolumn{2}{c}{KL-F4} &\multicolumn{2}{c}{KL-F8} & \multicolumn{2}{c}{KL-F4} &\multicolumn{2}{c}{KL-F8} & \multicolumn{2}{c}{KL-F4} \\
            \cmidrule(lr){2-3}\cmidrule(lr){4-5}\cmidrule(lr){6-7}\cmidrule(lr){8-9}\cmidrule(lr){10-11}\cmidrule(lr){12-13}\cmidrule(lr){14-15}\cmidrule(lr){16-17}
            Method & FID & Acc & FID & Acc & FID & Acc & FID & Acc & FID & Acc & FID & Acc & FID & Acc & FID & Acc \\ \midrule
            Pretraining & 9.97 & - & 3.29 & - & 11.10 & - & 5.65 & - & 9.44 & - & 3.26 & - & 8.12 & - & 2.70 & - \\ \midrule
            DPDM\textsubscript{CFG-1.0} &  25.02 & 52.03 & 12.94 & 57.24 & 26.52 & 49.03 & 23.31 & 53.37 & 21.63 & 64.22 & 18.16 & 59.35 & 14.64 & 75.23 & 11.37 & 76.30 \\
            DPDM\textsubscript{CFG-3.0} &  97.84 & - & 211.50 & - & 101.90 & - & 139.60 & - & 67.91 & - & 122.73 & - & 49.28 & - & 12.63 & - \\
            DPDM\textsubscript{per-label} &  \textbf{12.46} & \textbf{70.54} & \textbf{4.62} & \textbf{73.02} & \textbf{14.85} & \textbf{64.57} &\textbf{ 8.14 }& \textbf{68.33} & \textbf{13.72} & \textbf{71.30} & \textbf{7.38} & \textbf{76.34} & \textbf{13.26} & \textbf{77.29} & \textbf{6.42} & \textbf{77.73} \\
            \midrule
            DP-Promise\textsubscript{CFG-1.0} &  18.54 & 54.98 & \textbf{7.46} & 48.33 & 23.77 & 51.33 & \textbf{36.47} & 51.99 & \textbf{17.69} & 59.14 & \textbf{35.98} & 36.76 & 17.71 & 76.27 & \textbf{6.04} & 77.05 \\
            DP-Promise\textsubscript{CFG-3.0} &  85.64 & - & 121.90 & - & 70.20 & - & 70.54 & - & 50.90 & - & 151.16 & - & 53.91 & - & 15.30 & - \\
            DP-Promise\textsubscript{per-label} &  \textbf{12.35} & \textbf{72.05} & 15.56 & \textbf{70.80} & \textbf{15.63} & \textbf{67.37} & 43.77 & \textbf{67.59} & 18.80 & \textbf{73.14} & 71.06 & \textbf{71.50} & \textbf{13.20} & \textbf{77.70} & 6.98 & \textbf{78.44} \\
            \bottomrule
        \end{tabular}
    }
    \caption{
        Performance comparison of various conditioning, pretraining, and private training setups on the \rvlcdip{} dataset.
        The FID is computed between 50K real and generated samples, and the accuracy is obtained for each approach by training a ConvNeXt-B model on 50K generated samples. Layout+Class Cond., along with per-label private fine-tuning, shows the best downstream performance across all setups.
    }
    \label{table:comparison}
    \vspace{-2em}
\end{table}
\begin{figure}[!t]
\centering
\subfloat[$\varepsilon=1$]{%
    \includegraphics[width=\linewidth]{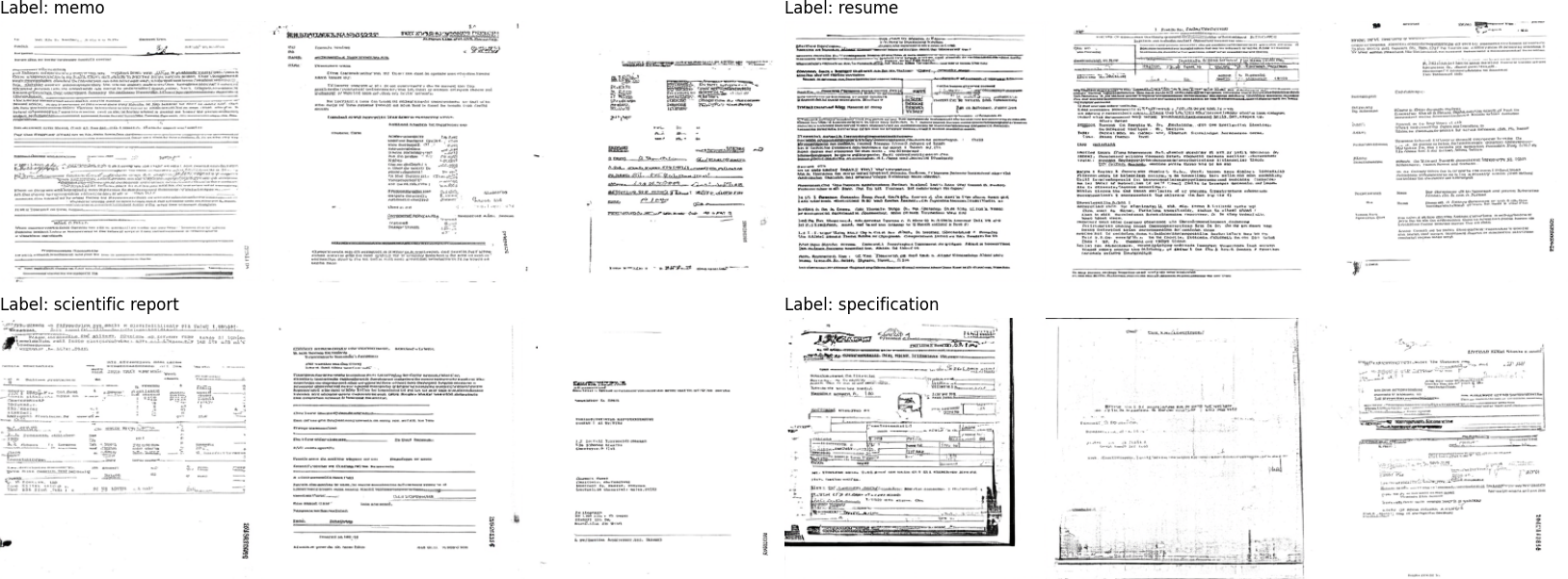}
    \label{fig:rvl-cdip-eps-1}
}
\hfill
\subfloat[$\varepsilon=5$]{%
    \includegraphics[width=\linewidth]{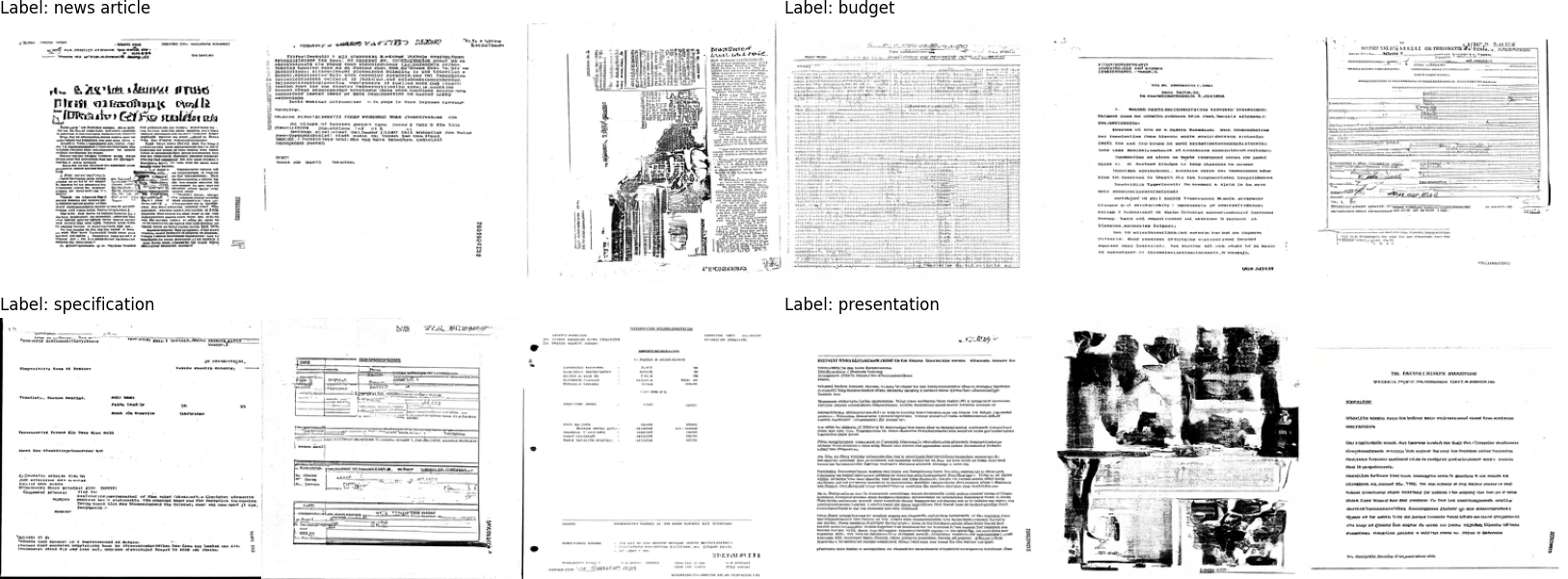}
    \label{fig:rvl-cdip-eps-2}
}
\hfill
\subfloat[$\varepsilon=10$]{%
    \includegraphics[width=\linewidth]{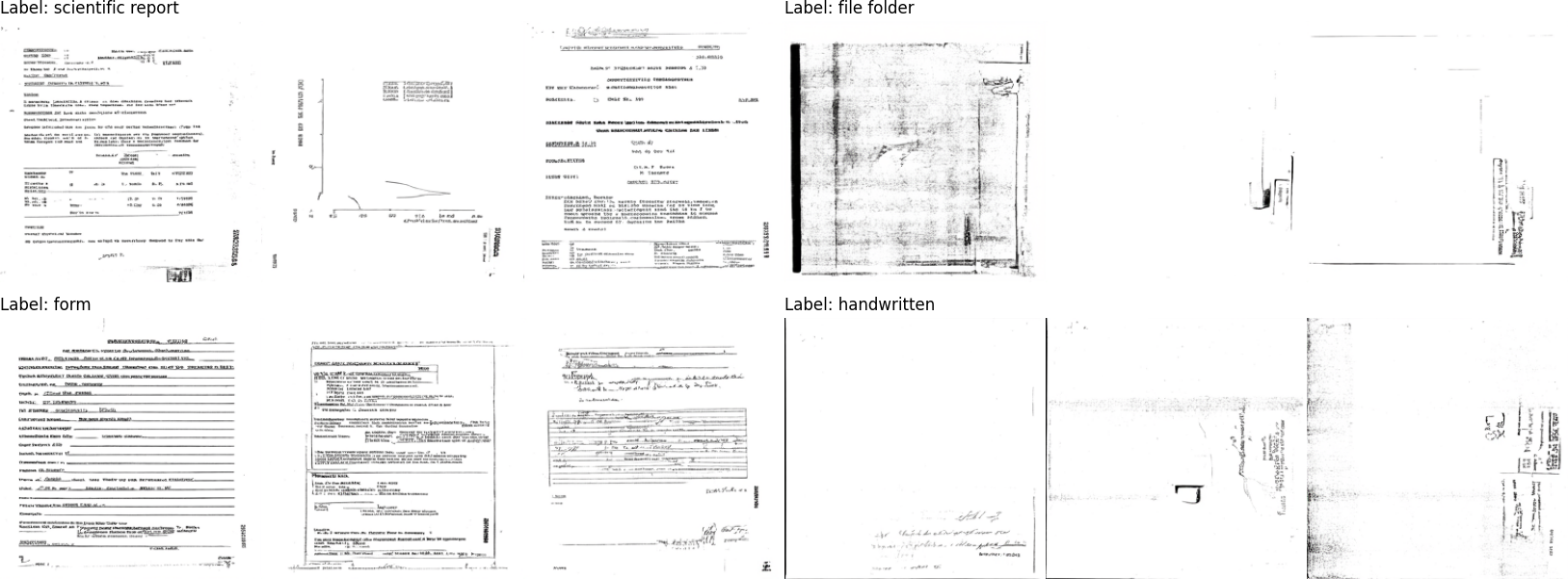}
    \label{fig:rvl-cdip-eps-3}
}
\caption{
    Visual comparison of the synthetic \rvlcdip{} samples generated using KL-F4 autoencoder at $\varepsilon \in \{1, 5, 10\}$. The visual quality of samples remained comparable between $\varepsilon = 1$ and $\varepsilon = 10$.
}
\label{fig:qualitative-1}
\vspace{-2em}
\end{figure}

Table~\ref{table:comparison} provides a comparison of our approach across different conditioning and private training setups.
By comparing the performance between the {Uncond.+Unaug.} and {Uncond.+Aug.} pretraining setups, we observe slight performance drops on {Uncond.+Aug.} compared to {Uncond.+Unaug.} case.
These performance drops are expected, as private fine-tuning is tasked with generating full documents, while the {Uncond.+Aug.} setup uses an augmented pretraining dataset with randomly cropped document images.
Nevertheless, this demonstrates that even with a significantly different pretraining distribution, the model can learn to generate complete document layouts with reasonable accuracy, especially in per-label training setups, where it can achieve a downstream accuracy of up to 68.33\% with the KL-F4 autoencoder.
On the other hand, if we compare the performances across different private training strategies (DPDM~\cite{dockhorn2023differentially} and DP-Promise~\cite{dppromise}), we find that while the model achieves sufficiently good FID scores with class-conditional private fine-tuning (DPDM\textsubscript{CFG-1.0} and DP-Promise\textsubscript{CFG-1.0} cases), it struggles to effectively distinguish between different classes, leading to suboptimal downstream performance. Moreover, we observe that using classifier-free guidance (DPDM\textsubscript{CFG-3.0} and DP-Promise\textsubscript{CFG-3.0} cases) with a guidance scale of 3.0 severely degrades the generative performance.
In contrast to class-conditional private training, training an independent model for each class (DPDM\textsubscript{per-label} and DP-Promise\textsubscript{per-label} cases) while maintaining the same privacy guarantees results in significant performance improvements.

With the introduction of class conditioning during pretraining (Class Cond. case), we observe minor performance improvements in downstream evaluation, suggesting that the additional class information was slightly beneficial to the model.
However, even with additional class information in pretraining, the class-conditional private fine-tuning (DPDM\textsubscript{CFG-1.0} and DP-Promise\textsubscript{CFG-1.0}) performs poorly on downstream evaluation.
In contrast, with additional layout information introduced in both pretraining and private fine-tuning (Layout + Class Cond. case), the model not only achieves significant performance improvements in DPDM\textsubscript{per-label} and DP-Promise\textsubscript{per-label} setups but also gains significant improvements on the class-conditional private fine-tuning setups (DPDM\textsubscript{CFG-1.0} and DP-Promise\textsubscript{CFG-1.0}). Overall, the Layout+Class Cond. pretraining setup achieved the best downstream performance (with accuracies of 78.44\% and 77.73\% on DPDM\textsubscript{per-label} and DP-Promise\textsubscript{per-label} settings, respectively), indicating the effectiveness of layout conditioning in improving the diversity of generated samples.

Note that while DP-Promise\textsubscript{per-label} with Layout+Class Cond. setup achieved the best performance, it resulted in significantly higher FID scores for the KL-F4 autoencoder under other pretraining setups.
Since DP-Promise~\cite{dppromise} applies local differential privacy by introducing noise into the samples, the amount of noise added depends on the embedding size.
Specifically, with an embedding size of $64 \times 64 \times 3$ in the KL-F4 autoencoder, the noise required to achieve a given privacy level was considerably higher than that in the KL-F8.
In addition, DP-Promise~\cite{dppromise} required manual tuning of the noise multiplier $\sigma$ for each scenario.
Therefore, for simplicity and to avoid excessive noise addition in the KL-F4 case (especially at higher privacy levels, such as $\varepsilon=1$ in the final evaluation), we opted to use the DPDM\textsubscript{per-label} training strategy with Layout+Class Cond. pretraining setup for all final downstream evaluations.

\begin{figure}[!t]
    \centering
    \subfloat[KL-F8]{%
        \includegraphics[width=\linewidth]{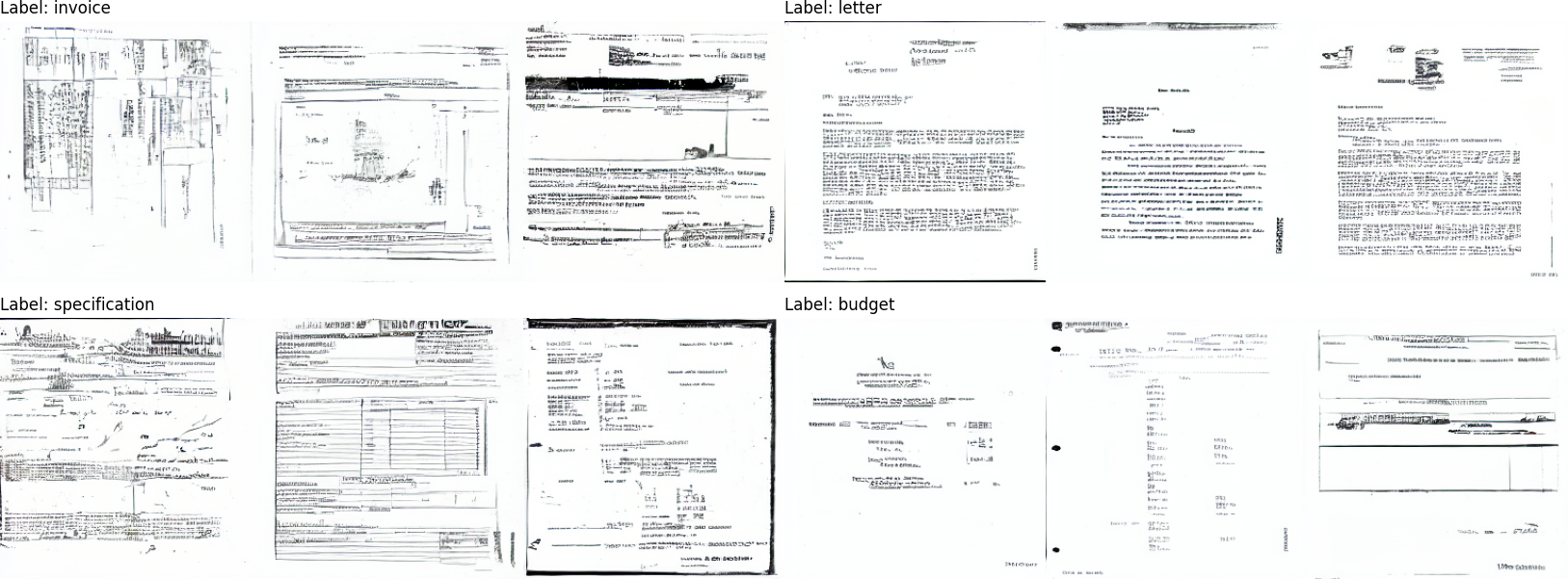}
        \label{fig:rvl-cdip-eps-1-f4-vs-f8-1}
    }
    \hfill
    \subfloat[KL-F4]{%
        \includegraphics[width=\linewidth]{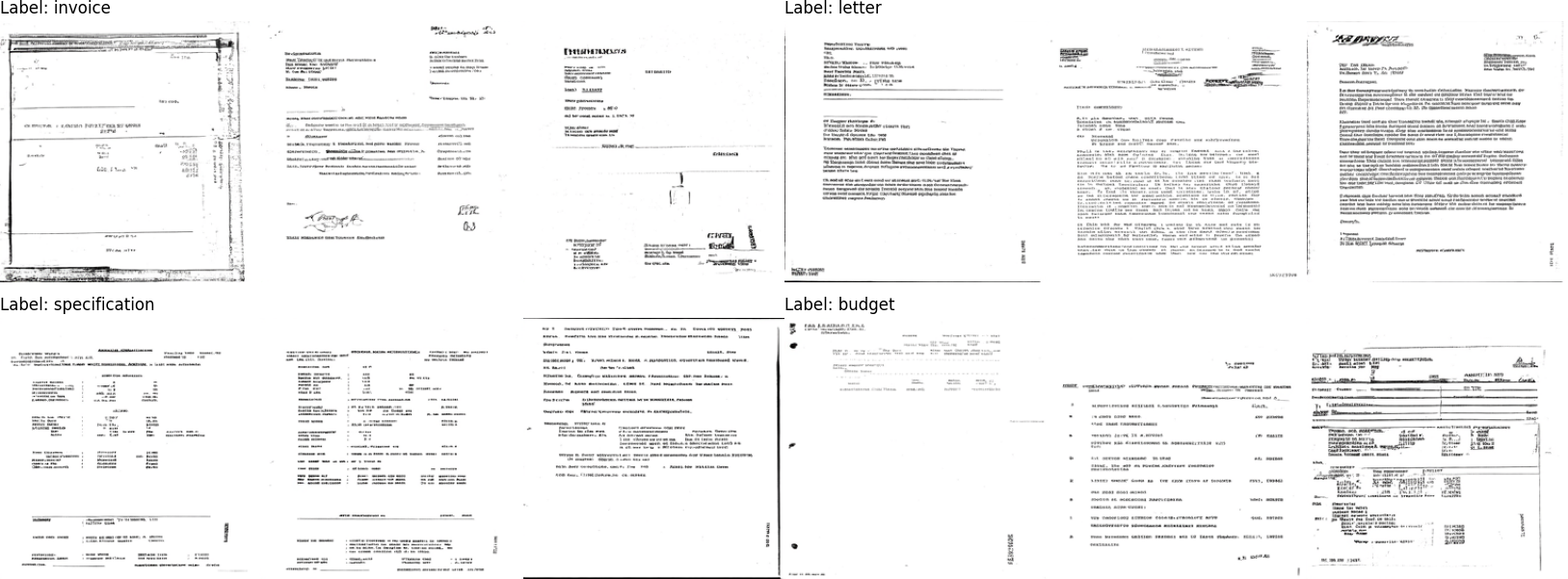}
        \label{fig:rvl-cdip-eps--2f4-vs-f8-1}
    }
    \caption{
        Visual comparison of synthetic \rvlcdip{} samples generated using KL-F8 and KL-F4 autoencoders at $\varepsilon = 1$. The sample quality with the KL-F4 autoencoder is significantly better across various document classes.
    }
    \label{fig:qualitative-2}
    \vspace{-1em}
\end{figure}

\subsection{Stage 2: Comprehensive Final Evaluation}
\subsubsection{Qualitative Evaluation}
In Fig.~\ref{fig:qualitative-1} and Fig.~\ref{fig:qualitative-2} we present document samples generated using the DPDM\textsubscript{per-label} (with Layout+Class Cond.) training strategy, across varying privacy setups $\varepsilon \in \{1, 5, 10\}$.
From Fig.~\ref{fig:qualitative-1}, we observe that our proposed approach is capable of generating a diverse set of document layouts across various document classes and privacy levels.
Furthermore, we find that the visual quality of samples generated with $\varepsilon = 10$ remained comparable to those generated with $\varepsilon = 1$.
In Fig.~\ref{fig:qualitative-2}, we also present a visual comparison between the samples generated using KL-F8 and KL-F4 autoencoders for a random subset of \rvlcdip{} classes.
As expected, the samples generated with KL-F4 exhibited much better quality and were closer to the real data distribution compared to those generated with KL-F8. Additional qualitative results for the samples generated on the Tobacco3482 dataset are provided in Appendix~\ref{app:qual-tobacco3482}.

\subsection{Quantitative Results}
\begin{table}[t]
    \centering
    \setlength{\tabcolsep}{0.5em}
    \resizebox{0.75\textwidth}{!}{
    \begin{tabular}{lcccccccc}
        \toprule
        & \multicolumn{2}{c}{$\varepsilon=1$} & \multicolumn{2}{c}{$\varepsilon=5$} & \multicolumn{2}{c}{$\varepsilon=10$}\\ \cmidrule(lr){2-3} \cmidrule(lr){4-5} \cmidrule(lr){6-7}
        Dataset& KL-F8 & KL-F4 &  KL-F8 & KL-F4 & KL-F8 & KL-F4 \\
        \midrule
        RVL-CDIP & 15.82 & 6.59 & 12.68 & 4.78 & 11.98 & 4.69 \\
        Tobacco3482 & 45.52 & 40.75 & 28.58 & 19.07 & 26.83 & 16.14\\
        \bottomrule
    \end{tabular}
}
    \caption{FID scores achieved using the DPDM\textsubscript{per-label} (with Layout+Class Cond.) training strategy under varying levels of privacy $\varepsilon \in \{1, 5, 10\}$. The FID is computed between 50K real and generated samples for each setup on the RVL-CDIP dataset and 1K real and generated samples for the Tobacco3482 dataset. On both datasets, KL-F4 leads to significantly better sample quality compared to KL-F8.
    }
    \label{table:final-fid}
    \vspace{-2em}
\end{table}

\begin{table}[t]
    \centering
    \setlength{\tabcolsep}{0.5em}
    \resizebox{\textwidth}{!}{
        \begin{tabular}{lllcccc}
            \toprule
            Dataset&Model&VAE& Baseline Acc.  & $\varepsilon=1$&$\varepsilon=5$&$\varepsilon=10$\\
            \midrule
            \multirow{8}{*}{RVL-CDIP}&
            ResNet-50/224\textsubscript{DP-SGD}~\cite{Saifullah2024}&-&90.50&-&78.34&79.21\\
            & ConvNeXt-B/384\textsubscript{DP-SGD}~\cite{Saifullah2024}&-&93.64&-&79.68&82.32\\
            \cmidrule(lr){2-7}
            \cmidrule(lr){2-7}
            & ResNet-50/224\textsubscript{DP-DocLDM} (ours)&KL-F8&90.50&74.31&76.95&78.15\\
            & ConvNeXt-B/224\textsubscript{DP-DocLDM} (ours)&KL-F8&92.86&78.72&82.51&\textbf{82.62}\\
            & DiT-B/224\textsubscript{DP-DocLDM} (ours)&KL-F8&92.89&77.89&81.91&82.38\\
            \cmidrule(lr){2-7}
            \cmidrule(lr){2-7}
            & ResNet-50/224\textsubscript{DP-DocLDM} (ours)&KL-F4 &90.50&74.74&78.24&73.23\\
            & ConvNeXt-B/224\textsubscript{DP-DocLDM} (ours)&KL-F4 &92.86&77.85&81.67&78.58\\
            & DiT-B/224\textsubscript{DP-DocLDM} (ours)&KL-F4&92.89&\textbf{80.21}&\textbf{83.18}\textbf{}&79.54\\
            \midrule
            \multirow{8}{*}{Tobacco3482} &
            ResNet-50/224\textsubscript{DP-SGD}~\cite{Saifullah2024}&-&82.42&-&44.44&46.29\\
            & ConvNeXt-B/384\textsubscript{DP-SGD}~\cite{Saifullah2024}&-&89.42&-&71.58&74.29\\
            \cmidrule(lr){2-7}
            \cmidrule(lr){2-7}
            & ResNet-50/224\textsubscript{DP-DocLDM} (ours)&KL-F8&82.42&65.86&	76.85&	76.00\\
            & ConvNeXt-B/224\textsubscript{DP-DocLDM} (ours)&KL-F8&89.42&70.50&	76.00&	80.43\\
            & DiT-B/224\textsubscript{DP-DocLDM} (ours)&KL-F8&92.71&69.85&	79.57&	80.14\\
            \cmidrule(lr){2-7}
            \cmidrule(lr){2-7}
            & ResNet-50/224\textsubscript{DP-DocLDM} (ours)&KL-F4&82.42&65.14&	77.71&	78.28\\
            & ConvNeXt-B/224\textsubscript{DP-DocLDM} (ours)&KL-F4&89.42&\textbf{72.57}&	\textbf{80.00}&	\textbf{82.57}\\
            & DiT-B/224\textsubscript{DP-DocLDM} (ours)&KL-F4&92.71&67.28&	79.57&	\textbf{82.57}\\
            \bottomrule
        \end{tabular}
    }
    \caption{Downstream evaluation results for the proposed DPDM\textsubscript{per-label} (with Layout+Class Cond.) training strategy on \rvlcdip{} and Tobacco3482. Our approach brings minor performance improvements on the large-scale \rvlcdip{} dataset and significant improvements on the small-scale Tobacco3482 dataset.
    }
    \vspace{-3em}
    \label{table:final-acc}
\end{table}
In Table~\ref{table:final-fid}, we present the FID scores achieved using the DPDM\textsubscript{per-label} (with Layout+Class Cond.) training strategy under varying levels of privacy $\varepsilon \in \{1, 5, 10\}$.
We observe that on the \rvlcdip{} dataset, the KL-F4 autoencoder demonstrated significantly better FID scores compared to the KL-F8 autoencoder, with only minor drops in performance when increasing the privacy level from $\varepsilon=10$ to $\varepsilon=1$. This supports the qualitative results presented in the previous section.
We observe a similar pattern for the Tobacco3482 dataset; however, the overall FID scores were much higher for $\varepsilon=1$ compared to $\varepsilon=10$.

In Table~\ref{table:final-acc}, we present the downstream evaluation results for the proposed DPDM\textsubscript{per-label} (with Layout+Class Cond.) training strategy and compare its performance with previous work~\cite{Saifullah2024} that utilizes DP-Adam~\cite{dpsgd-Abadi2016} to directly train the classifiers under private settings.
Examining the downstream performance on the \rvlcdip{} dataset, we observe that ResNet-50~\cite{resnet} with our proposed approach slightly underperformed compared to direct training under DP-Adam~\cite{dpsgd-Abadi2016}.
However, with ConvNeXt-B~\cite{liu2022convnet} we observed slight performance improvements in both $\varepsilon=5$ and $\varepsilon=10$ settings with just a $224\times224$ resolution compared to $384\times384$ resolution used in previous work~\cite{Saifullah2024}.
Surprisingly, while KL-F4 resulted in better visual sample quality, it led to significantly poorer downstream performance compared to KL-F8 on $\varepsilon=10$.
On $\varepsilon=5$, however, both KL-F8 and KL-F4 led to better downstream performance compared to DP-Adam~\cite{dpsgd-Abadi2016}. Lastly, the DiT-B~\cite{doc-vit} model, with its self-supervised document-specific pretraining, outperformed both ResNet-50~\cite{resnet} and ConvNeXt-B~\cite{liu2022convnet} across all privacy settings,
achieving a reasonable downstream performance of 80.21\% at $\varepsilon=1$.
Overall, with our proposed approach, we achieved slight performance improvements on the \rvlcdip{} dataset compared to existing approaches, with comparable performance observed across low ($\varepsilon=10$) and high privacy regimes ($\varepsilon=1$)

The effectiveness of our proposed approach is more clearly highlighted on the small-scale Tobacco3482 dataset, where DP-Adam~\cite{dpsgd-Abadi2016} leads to significant performance drops for both ResNet-50~\cite{resnet} (with accuracies of 44\%-46\%) and ConvNeXt-B~\cite{liu2022convnet} (with accuracies of 71\%-74\%) models.
In contrast, with our proposed approach, both models were able to achieve downstream accuracies of 78\%-82\%, while also demonstrating reasonable performance under $\varepsilon=1$.
Overall, these results suggest that our proposed method could serve as a more viable alternative to DP-Adam~\cite{dpsgd-Abadi2016} for private training, particularly for small-scale document datasets.

\section{Conclusion}
In this work, we introduced a novel approach for private synthetic document image generation and demonstrated its effectiveness (especially on small datasets) through downstream evaluation on document classification. However, there are some limitations in our approach that are worth discussing.
Our approach requires multiple steps for evaluation, making it costly to explore a broader range of hyperparameter configurations for private training, which is often necessary for the DP training. In this work, we used a fixed set of parameters for all our experiments; however, better configurations may exist. Therefore, future research could investigate more recent hyperparameter-free variants~\cite{liu2025towards,NEURIPS2021_91cff01a} of DP-Adam~\cite{dpsgd-Abadi2016} to eliminate the need for parameter tuning.
Additionally, for generating synthetic datasets, we used the standard DDPM sampler~\cite{ho2020denoisingdiffusionprobabilisticmodels} in this work. In the future, it will be worthwhile to explore more complex sampling approaches, such as CADS~\cite{sadat2024cadsunleashingdiversitydiffusion}, to improve sample diversity and reduce overfitting in downstream training.
Lastly, in this work, we only explored the potential of private document generation for the document classification task. In the future, it will be worthwhile to develop similar approaches for introducing privacy in more complex document analysis tasks, such as layout analysis, handwriting recognition, and table extraction.
\bibliographystyle{splncs04}

\begin{thebibliography}{10}
\providecommand{\url}[1]{\texttt{#1}}
\providecommand{\urlprefix}{URL }
\providecommand{\doi}[1]{https://doi.org/#1}

\bibitem{dpsgd-Abadi2016}
Abadi, M., Chu, A., Goodfellow, I., McMahan, H.B., Mironov, I., Talwar, K.,
Zhang, L.: Deep learning with differential privacy. In: Proceedings of the
2016 {ACM} {SIGSAC} Conference on Computer and Communications Security. {ACM}
(oct 2016), \url{https://doi.org/10.1145%2F2976749.2978318}
	
	\bibitem{Afzal2017-deep-cnns}
	Afzal, M.Z., Kolsch, A., Ahmed, S., Liwicki, M.: Cutting the error by half:
	Investigation of very deep cnn and advanced training strategies for document
	image classification. Proceedings of the International Conference on Document
	Analysis and Recognition, ICDAR  \textbf{1},  883--888 (2017).
	\doi{10.1109/ICDAR.2017.149}
	
	\bibitem{attacks-survey-Al-Rubaie2019}
	Al-Rubaie, M., Chang, J.M.: {Privacy-Preserving Machine Learning: Threats and
		Solutions}. IEEE Secur. Priv.  \textbf{17}(2),  49--58 (mar 2019)
	
	\bibitem{NEURIPS2021_91cff01a}
	Andrew, G., Thakkar, O., McMahan, B., Ramaswamy, S.: Differentially private
	learning with adaptive clipping. In: Ranzato, M., Beygelzimer, A., Dauphin,
	Y., Liang, P., Vaughan, J.W. (eds.) Advances in Neural Information Processing
	Systems. vol.~34, pp. 17455--17466. Curran Associates, Inc. (2021),
	\url{https://proceedings.neurips.cc/paper_files/paper/2021/file/91cff01af640a24e7f9f7a5ab407889f-Paper.pdf}
	
	\bibitem{priv-documents-Basu2021}
	Basu, P., Roy, T.S., Naidu, R., Muftuoglu, Z.: {Privacy enabled Financial Text
		Classification using Differential Privacy and Federated Learning}. In:
	Proceedings of the 3rd Workshop on Economics and Natural Language Processing,
	ECONLP 2021. pp. 50--55. Association for Computational Linguistics,
	Stroudsburg, PA, USA (oct 2021),
	\url{https://aclanthology.org/2021.econlp-1.7}
	
	\bibitem{dp-basu2022benchmarking}
	Basu, P., Roy, T.S., Naidu, R., Muftuoglu, Z., Singh, S., Mireshghallah, F.:
	Benchmarking differential privacy and federated learning for bert models
	(2022)
	
	\bibitem{memorization-Carlini2019}
	Carlini, N., Liu, C., Erlingsson, {\'{U}}., Kos, J., Song, D.: {The secret
		Sharer: Evaluating and testing unintended memorization in neural networks}.
	In: Proc. 28th USENIX Secur. Symp. pp. 267--284 (2019)
	
	\bibitem{text_from_llm-carlini}
	Carlini, N., Tram{\`e}r, F., Wallace, E., Jagielski, M., Herbert-Voss, A., Lee,
	K., Roberts, A., Brown, T., Song, D., Erlingsson, {\'U}., Oprea, A., Raffel,
	C.: Extracting training data from large language models. In: 30th USENIX
	Security Symposium (USENIX Security 21). pp. 2633--2650. USENIX Association
	(aug 2021),
	\url{https://www.usenix.org/conference/usenixsecurity21/presentation/carlini-extracting}
	
	\bibitem{NEURIPS2020_9547ad6b}
	Chen, D., Orekondy, T., Fritz, M.: Gs-wgan: A gradient-sanitized approach for
	learning differentially private generators. In: Larochelle, H., Ranzato, M.,
	Hadsell, R., Balcan, M., Lin, H. (eds.) Advances in Neural Information
	Processing Systems. vol.~33, pp. 12673--12684. Curran Associates, Inc.
	(2020),
	\url{https://proceedings.neurips.cc/paper_files/paper/2020/file/9547ad6b651e2087bac67651aa92cd0d-Paper.pdf}
	
	\bibitem{model-inversion-Coavoux2020}
	Coavoux, M., Narayan, S., Cohen, S.B.: {Privacy-preserving neural
		representations of text}. In: Proc. 2018 Conf. Empir. Methods Nat. Lang.
	Process. EMNLP 2018. pp. 1--10 (2020). \doi{10.18653/v1/d18-1001}
	
	\bibitem{bert-devlin2019}
	Devlin, J., Chang, M.W., Lee, K., Toutanova, K.: {BERT}: Pre-training of deep
	bidirectional transformers for language understanding. In: Proceedings of the
	2019 Conference of the North {A}merican Chapter of the Association for
	Computational Linguistics: Human Language Technologies, Volume 1 (Long and
	Short Papers). pp. 4171--4186. Association for Computational Linguistics,
	Minneapolis, Minnesota (jun 2019), \url{https://aclanthology.org/N19-1423}
	
	\bibitem{dhariwal2021diffusionmodelsbeatgans}
	Dhariwal, P., Nichol, A.: Diffusion models beat gans on image synthesis (2021),
	\url{https://arxiv.org/abs/2105.05233}
	
	\bibitem{dockhorn2023differentially}
	Dockhorn, T., Cao, T., Vahdat, A., Kreis, K.: Differentially private diffusion
	models (2023), \url{https://openreview.net/forum?id=pX21pH4CsNB}
	
	\bibitem{dosovitskiy2021an}
	Dosovitskiy, A., Beyer, L., Kolesnikov, A., Weissenborn, D., Zhai, X.,
	Unterthiner, T., Dehghani, M., Minderer, M., Heigold, G., Gelly, S.,
	Uszkoreit, J., Houlsby, N.: An image is worth 16x16 words: Transformers for
	image recognition at scale. In: International Conference on Learning
	Representations (2021), \url{https://openreview.net/forum?id=YicbFdNTTy}
	
	\bibitem{dp-Dwork2014}
	Dwork, C., Roth, A., Dwork, C., Roth, A.: {The Algorithmic Foundations of
		Differential Privacy}. Foundations and Trends R in Theoretical Computer
	Science  \textbf{9},  211--407 (2014). \doi{10.1561/0400000042}
	
	\bibitem{dp-Dwork2006}
	Dwork, C.: {Differential Privacy}. In: Automata, Languages and Programming. pp.
	1--12. Springer Berlin Heidelberg, Berlin, Heidelberg (2006),
	\url{http://link.springer.com/10.1007/11787006{\_}1}
	
	\bibitem{GDPR2016a}
	{European Parliament}, {Council of the European Union}: Regulation ({EU})
	2016/679 of the {European} {Parliament} and of the {Council},
	\url{https://data.europa.eu/eli/reg/2016/679/oj}
	
	\bibitem{ferrando2020-doc-class-4}
	Ferrando, J., Dom{\'i}nguez, J.L., Torres, J., Garc{\'i}a, R., Garc{\'i}a, D.,
	Garrido, D., Cortada, J., Valero, M.: Improving accuracy and speeding up
	document image classification through parallel systems. In: Krzhizhanovskaya,
	V.V., Z{\'a}vodszky, G., Lees, M.H., Dongarra, J.J., Sloot, P.M.A., Brissos,
	S., Teixeira, J. (eds.) Computational Science -- ICCS 2020. pp. 387--400.
	Springer International Publishing, Cham (2020)
	
	\bibitem{local-mdp-Feyisetan2019}
	Feyisetan, O., Diethe, T., Drake, T.: {Leveraging hierarchical representations
		for preserving privacy and utility in text}. In: Proceedings - IEEE
	International Conference on Data Mining, ICDM. vol. 2019-Novem, pp. 210--219
	(oct 2019), \url{http://arxiv.org/abs/1910.08917}
	
	\bibitem{model-inversion-att-Fredrikson2015}
	Fredrikson, M., Jha, S., Ristenpart, T.: {Model inversion attacks that exploit
		confidence information and basic countermeasures}. In: Proc. ACM Conf.
	Comput. Commun. Secur. vol. 2015-Octob, pp. 1322--1333. ACM, New York, NY,
	USA (2015), \url{http://dx.doi.org/10.1145/2810103.2813677}
	
	\bibitem{ghalebikesabi2023differentiallyprivatediffusionmodels}
	Ghalebikesabi, S., Berrada, L., Gowal, S., Ktena, I., Stanforth, R., Hayes, J.,
	De, S., Smith, S.L., Wiles, O., Balle, B.: Differentially private diffusion
	models generate useful synthetic images (2023),
	\url{https://arxiv.org/abs/2302.13861}
	
	\bibitem{guan2024idnetnoveldatasetidentity}
	Guan, H., Wang, Y., Xie, L., Nag, S., Goel, R., Swamy, N.E.N., Yang, Y., Xiao,
	C., Prisby, J., Maciejewski, R., Zou, J.: Idnet: A novel dataset for identity
	document analysis and fraud detection (2024),
	\url{https://arxiv.org/abs/2408.01690}
	
	\bibitem{hamdani-doc}
	Hamdani, S.J.H., Saifullah, S., Agne, S., Dengel, A., Ahmed, S.: Latent
	diffusion for guided document table generation. In: Barney~Smith, E.H.,
	Liwicki, M., Peng, L. (eds.) Document Analysis and Recognition - ICDAR 2024.
	pp. 368--383. Springer Nature Switzerland, Cham (2024)
	
	\bibitem{harder2023pretrained}
	Harder, F., Jalali, M., Sutherland, D.J., Park, M.: Pre-trained perceptual
	features improve differentially private image generation. Transactions on
	Machine Learning Research  (2023),
	\url{https://openreview.net/forum?id=R6W7zkMz0P}
	
	\bibitem{harley2015icdar}
	Harley, A.W., Ufkes, A., Derpanis, K.G.: Evaluation of deep convolutional nets
	for document image classification and retrieval. In: International Conference
	on Document Analysis and Recognition ({ICDAR})
	
	\bibitem{resnet}
	He, K., Zhang, X., Ren, S., Sun, J.: Deep residual learning for image
	recognition. 2016 IEEE Conference on Computer Vision and Pattern Recognition
	(CVPR) pp. 770--778 (2015),
	\url{https://api.semanticscholar.org/CorpusID:206594692}
	
	\bibitem{diffusion-layout-doc}
	He, L., Lu, Y., Corring, J., Florencio, D., Zhang, C.: Diffusion-Based Document
	Layout Generation, p. 361–378. Springer Nature Switzerland (2023),
	\url{http://dx.doi.org/10.1007/978-3-031-41676-7_21}
	
	\bibitem{ho2020denoisingdiffusionprobabilisticmodels}
	Ho, J., Jain, A., Abbeel, P.: Denoising diffusion probabilistic models (2020),
	\url{https://arxiv.org/abs/2006.11239}
	
	\bibitem{ho2022classifierfreediffusionguidance}
	Ho, J., Salimans, T.: Classifier-free diffusion guidance (2022),
	\url{https://arxiv.org/abs/2207.12598}
	
	\bibitem{dp-Hoory2021}
	Hoory, S., Feder, A., Tendler, A., Cohen, A., Erell, S., Laish, I., Nakhost,
	H., Stemmer, U., Benjamini, A., Hassidim, A., Matias, Y.: {Learning and
		Evaluating a Differentially Private Pre-trained Language Model}. In: Findings
	of the Association for Computational Linguistics, Findings of ACL: EMNLP
	2021. pp. 1178--1189 (2021). \doi{10.18653/v1/2021.privatenlp-1.3}
	
	\bibitem{priv-survey-Hu2023}
	Hu, L., Habernal, I., Shen, L., Wang, D.: Differentially private natural
	language models: Recent advances and future directions. arXiv
	\textbf{abs/2301.09112} (2023), \url{https://arxiv.org/abs/2301.09112}
	
	\bibitem{layoutlmv3-Huang2022}
	Huang, Y., Lv, T., Cui, L., Lu, Y., Wei, F.: {LayoutLMv3: Pre-training for
		Document AI with Unified Text and Image Masking}. In: Proceedings of the 30th
	ACM International Conference on Multimedia. pp. 4083--4091. ACM, New York,
	NY, USA (oct 2022), \url{https://dl.acm.org/doi/10.1145/3503161.3548112}
	
	\bibitem{dp-gdp-koskela2022individual}
	Koskela, A., Tobaben, M., Honkela, A.: Individual privacy accounting with
	gaussian differential privacy. In: The Eleventh International Conference on
	Learning Representations (2023),
	\url{https://openreview.net/forum?id=JmC_Tld3v-f}
	
	\bibitem{pixeldp}
	Lecuyer, M., Atlidakis, V., Geambasu, R., Hsu, D., Jana, S.: Certified
	robustness to adversarial examples with differential privacy (2019),
	\url{https://arxiv.org/abs/1802.03471}
	
	\bibitem{formnet-Lee2022}
	Lee, C.Y., Li, C.L., Dozat, T., Perot, V., Su, G., Hua, N., Ainslie, J., Wang,
	R., Fujii, Y., Pfister, T.: {FormNet: Structural Encoding beyond Sequential
		Modeling in Form Document Information Extraction}. In: Proceedings of the
	Annual Meeting of the Association for Computational Linguistics. vol.~1, pp.
	3735--3754. Long Papers (2022). \doi{10.18653/v1/2022.acl-long.260}
	
	\bibitem{doc-vit}
	Li, J., Xu, Y., Lv, T., Cui, L., Zhang, C., Wei, F.: Dit: Self-supervised
	pre-training for document image transformer. In: Proceedings of the 30th ACM
	International Conference on Multimedia. p. 3530–3539. MM '22, Association
	for Computing Machinery, New York, NY, USA (2022),
	\url{https://doi.org/10.1145/3503161.3547911}
	
	\bibitem{dp-Li2021}
	Li, X., Tramer, F., Liang, P., Hashimoto, T.: Large language models can be
	strong differentially private learners. In: International Conference on
	Learning Representations (2022),
	\url{https://openreview.net/forum?id=bVuP3ltATMz}
	
	\bibitem{liew2022pearl}
	Liew, S.P., Takahashi, T., Ueno, M.: {PEARL}: Data synthesis via private
	embeddings and adversarial reconstruction learning. In: International
	Conference on Learning Representations (2022),
	\url{https://openreview.net/forum?id=M6M8BEmd6dq}
	
	\bibitem{liu2024differentially}
	Liu, M.F., Lyu, S., Vinaroz, M., Park, M.: Differentially private latent
	diffusion models. Transactions on Machine Learning Research  (2024),
	\url{https://openreview.net/forum?id=AkdQ266kHj}
	
	\bibitem{liu2025towards}
	Liu, R., Bu, Z.: Towards hyperparameter-free optimization with differential
	privacy. In: The Thirteenth International Conference on Learning
	Representations (2025), \url{https://openreview.net/forum?id=2kGKsyhtvh}
	
	\bibitem{liu2022convnet}
	Liu, Z., Mao, H., Wu, C.Y., Feichtenhofer, C., Darrell, T., Xie, S.: A convnet
	for the 2020s. Proceedings of the IEEE/CVF Conference on Computer Vision and
	Pattern Recognition (CVPR)  (2022)
	
	\bibitem{fl-rnn-McMahan2017LearningDP}
	McMahan, H.B., Ramage, D., Talwar, K., Zhang, L.: Learning differentially
	private recurrent language models. In: International Conference on Learning
	Representations (2017)
	
	\bibitem{fl-mercier2021evaluating}
	Mercier, D., Lucieri, A., Munir, M., Dengel, A., Ahmed, S.: Evaluating
	privacy-preserving machine learning in critical infrastructures: A case study
	on time-series classification. IEEE Transactions on Industrial Informatics
	(2021)
	
	\bibitem{dp-rdp-mironov}
	Mironov, I.: R{\'{e}}nyi differential privacy. In: 2017 {IEEE} 30th Computer
	Security Foundations Symposium ({CSF}). {IEEE} (aug 2017),
	\url{https://doi.org/10.1109%2Fcsf.2017.11}
		
		\bibitem{dp-papernot2020making}
		Papernot, N., Chien, S., Song, S., Thakurta, A., Erlingsson, U.: Making the
		shoe fit: Architectures, initializations, and tuning for learning with
		privacy (2020), \url{https://openreview.net/forum?id=rJg851rYwH}
		
		\bibitem{local-dp-cape-plant2021}
		Plant, R., Gkatzia, D., Giuffrida, V.: {CAPE}: Context-aware private embeddings
		for private language learning. In: Proceedings of the 2021 Conference on
		Empirical Methods in Natural Language Processing. pp. 7970--7978. Association
		for Computational Linguistics, Online and Punta Cana, Dominican Republic (nov
		2021), \url{https://aclanthology.org/2021.emnlp-main.628}
		
		\bibitem{tilt-Powalski2021}
		{Powalski Rafa{\l}and Borchmann}, {\L}., Jurkiewicz, D., Dwojak, T.,
		{Pietruszka Micha{\l}and Pa{\l}ka}, G.: {Going Full-TILT Boogie on Document
			Understanding with Text-Image-Layout Transformer}. In: Llad{\'{o}}s, J.,
		Lopresti, D., Uchida, S. (eds.) Document Analysis and Recognition -- ICDAR
		2021. pp. 732--747. Springer International Publishing, Cham (2021)
		
		\bibitem{rombach2021highresolution}
		Rombach, R., Blattmann, A., Lorenz, D., Esser, P., Ommer, B.: High-resolution
		image synthesis with latent diffusion models (2021)
		
		\bibitem{sadat2024cadsunleashingdiversitydiffusion}
		Sadat, S., Buhmann, J., Bradley, D., Hilliges, O., Weber, R.M.: Cads:
		Unleashing the diversity of diffusion models through condition-annealed
		sampling (2024), \url{https://arxiv.org/abs/2310.17347}
		
		\bibitem{Saifullah2022-docxclassifier}
		Saifullah, S., Agne, S., Dengel, A., Ahmed, S.: Docxclassifier: Towards an
		interpretable deep convolutional neural network for document image
		classification  (9 2022), \url{https://doi.org/10.36227/techrxiv.19310489.v4}
		
		\bibitem{dp-kie}
		Saifullah, S., Agne, S., Dengel, A., Ahmed, S.: Pried-kie: Towards privacy
		preserved document key information extraction (2023)
		
		\bibitem{Saifullah2024}
		Saifullah, S., Mercier, D., Agne, S., Dengel, A., Ahmed, S.: Towards privacy
		preserved document image classification: a comprehensive benchmark.
		International Journal on Document Analysis and Recognition (IJDAR)
		\textbf{27}(3),  475–499 (jun 2024),
		\url{http://dx.doi.org/10.1007/s10032-024-00469-8}
		
		\bibitem{layout-shen2021}
		Shen, Z., Zhang, R., Dell, M., Lee, B.C.G., Carlson, J., Li, W.: Layoutparser:
		A unified toolkit for deep learning based document image analysis. In:
		Llad{\'o}s, J., Lopresti, D., Uchida, S. (eds.) Document Analysis and
		Recognition -- ICDAR 2021. pp. 131--146. Springer International Publishing,
		Cham (2021)
		
		\bibitem{membership-inf-att-Shokri2017}
		Shokri, R., Stronati, M., Song, C., Shmatikov, V.: {Membership Inference
			Attacks Against Machine Learning Models}. In: Proc. - IEEE Symp. Secur. Priv.
		pp. 3--18 (2017). \doi{10.1109/SP.2017.41}
		
		\bibitem{TessOverview}
		Smith, R.: An overview of the tesseract ocr engine. In: ICDAR '07: Proceedings
		of the Ninth International Conference on Document Analysis and Recognition.
		pp. 629--633. IEEE Computer Society, Washington, DC, USA (2007),
		\url{https://storage.googleapis.com/pub-tools-public-publication-data/pdf/33418.pdf}
		
		\bibitem{iitcdip}
		Soboroff, I.: Complex document information processing (cdip) dataset, national
		institute of standards and technology (2022)
		
		\bibitem{sohldickstein2015deepunsupervisedlearningusing}
		Sohl-Dickstein, J., Weiss, E.A., Maheswaranathan, N., Ganguli, S.: Deep
		unsupervised learning using nonequilibrium thermodynamics (2015),
		\url{https://arxiv.org/abs/1503.03585}
		
		\bibitem{ldms-doc}
		Tanveer, N., Ul-Hasan, A., Shafait, F.: Diffusion models for document image
		generation. In: Document Analysis and Recognition - ICDAR 2023: 17th
		International Conference, San Jos\'{e}, CA, USA, August 21–26, 2023,
		Proceedings, Part III. p. 438–453. Springer-Verlag, Berlin, Heidelberg
		(2023), \url{https://doi.org/10.1007/978-3-031-41682-8_27}
		
		\bibitem{dp-docvqa}
		Tito, R., Nguyen, K., Tobaben, M., Kerkouche, R., Souibgui, M.A., Jung, K.,
		J{\"a}lk{\"o}, J., D'Andecy, V.P., Joseph, A., Kang, L., Valveny, E.,
		Honkela, A., Fritz, M., Karatzas, D.: Privacy-aware document visual question
		answering. In: Barney~Smith, E.H., Liwicki, M., Peng, L. (eds.) Document
		Analysis and Recognition - ICDAR 2024. pp. 199--218. Springer Nature
		Switzerland, Cham (2024)
		
		\bibitem{torkzadehmahani2019dp}
		Torkzadehmahani, R., Kairouz, P., Paten, B.: Dp-cgan: Differentially private
		synthetic data and label generation. In: Proceedings of the IEEE Conference
		on Computer Vision and Pattern Recognition Workshops. pp.~0--0 (2019)
		
		\bibitem{tsai2024differentiallyprivatefinetuningdiffusion}
		Tsai, Y.L., Li, Y., Chen, Z., Chen, P.Y., Yu, C.M., Ren, X., Buet-Golfouse, F.:
		Differentially private fine-tuning of diffusion models (2024),
		\url{https://arxiv.org/abs/2406.01355}
		
		\bibitem{dppromise}
		Wang, H., Pang, S., Lu, Z., Rao, Y., Zhou, Y., Xue, M.: dp-promise:
		Differentially private diffusion probabilistic models for image synthesis.
		In: USENIX Security Symposium (2024),
		\url{https://www.usenix.org/conference/usenixsecurity24/presentation/wang-haichen}
		
		\bibitem{dp-Wunderlich2022}
		Wunderlich, D., Bernau, D., Ald{\`{a}}, F., Parra-Arnau, J., Strufe, T.: {On
			the Privacy–Utility Trade-Off in Differentially Private Hierarchical Text
			Classification}. Applied Sciences (Switzerland)  \textbf{12}(21) (mar 2022),
		\url{http://arxiv.org/abs/2103.02895}
		
		\bibitem{image_from_grad-yin}
		Yin, H., Mallya, A., Vahdat, A., Alvarez, J.M., Kautz, J., Molchanov, P.: See
		through gradients: Image batch recovery via gradinversion. In: Proceedings of
		the IEEE/CVF Conference on Computer Vision and Pattern Recognition (CVPR).
		pp. 16337--16346 (June 2021)
		
		\bibitem{yun2019cutmixregularizationstrategytrain}
		Yun, S., Han, D., Oh, S.J., Chun, S., Choe, J., Yoo, Y.: Cutmix: Regularization
		strategy to train strong classifiers with localizable features (2019),
		\url{https://arxiv.org/abs/1905.04899}
		
		\bibitem{zhang2018mixupempiricalriskminimization}
		Zhang, H., Cisse, M., Dauphin, Y.N., Lopez-Paz, D.: mixup: Beyond empirical
		risk minimization (2018), \url{https://arxiv.org/abs/1710.09412}
		
		\bibitem{dp-medical}
		Ziller, A., Usynin, D., Braren, R., Makowski, M., Rueckert, D., Kaissis, G.:
		Medical imaging deep learning with differential privacy. Scientific Reports
		\textbf{11}(1) (jun 2021), \url{http://dx.doi.org/10.1038/s41598-021-93030-0}
		
	\end{thebibliography}

\newpage
\appendix
\section{Privacy Accounting}
\label{app:privacy-accounting}
To calculate the overall privacy loss $(\varepsilon)$ over a training cycle, we use the R\'enyi DP (RDP)~\cite{dp-rdp-mironov} privacy accountant for DPDM~\cite{dockhorn2023differentially} and Gaussian DP (GDP)~\cite{dp-gdp-koskela2022individual} for DP-Promise~\cite{dppromise}. Both RDP and GDP are improvements over the moments accountant proposed by Abadi~\etal~\cite{dpsgd-Abadi2016}, and can be used to track the overall release of information from a differentially private random algorithm. Given the data sampling rate $q$, a given noise multiplier $\sigma$ and target privacy budget $(\varepsilon,\delta)$, privacy accountants provide an estimate of the overall privacy loss over a fixed number of training steps. Inversely, standard numerical optimization can be applied to these accountants to estimate the required value of the noise multiplier $\sigma$ in order to achieve a target privacy loss $(\varepsilon,\delta)$ over the training cycle.  In this work, we perform this numerical optimization for all DPDM~\cite{dockhorn2023differentially} experiments to compute the required noise multiplier for a given epsilon $\varepsilon\in \{1,5,10\}$.
\section{Private Training Algorithm}
\label{app:privacy-algorithms}
For completeness, we provide the pseudo-code for the DPDM~\cite{dockhorn2023differentially} training algorithm, which is used in this work for private fine-tuning, in Algorithm~\ref{algo:dpdm}. DPDM~\cite{dockhorn2023differentially} is a variant of DP-Adam~\cite{dpsgd-Abadi2016} that uses noise multiplicity to improve the training of diffusion models~\cite{dockhorn2023differentially}. Note that we use the same DPDM~\cite{dockhorn2023differentially} variant in the private training step of DP-Promise~\cite{dppromise}.
\begin{algorithm}[H]
    \small
    \caption{{DPDM Training (DP-Adam with noise multiplicity $K$)}
    }
    \label{algo:dpdm}
    \begin{algorithmic}
        \State {\bfseries Input:} Private data set $d = \{x_{j}\}_{j=1}^N$, subsampling rate $B/N$, DP noise scale $\sigma_{\mathrm{DP}}$, clipping constant $C$, sampling function \textit{Poisson Sample}, denoiser $D_\mathrm{\theta}$ with initial parameters $\theta$, noise distribution $p(\sigma)$, learning rate $\eta$, total steps $T$, noise multiplicity $K$, Adam optimizer
        \State Output: Trained parameters $\theta$
        \For{$t=1$ {\bfseries to} $T$}
        \State $\mathcal{B} \sim \textit{Poisson Sample}(N, B/N)$
        \For{$i \in \mathcal{B}$}
        \State $\left\{(\sigma_{ik}, \mathrm{n}_{ik})\right \}_{k=1}^K \sim p(\sigma)\mathcal{N}(\mathbf{0},\sigma^2)$
        \State $\tilde{l}_{i} = \frac{1}{K} \sum_{k =1}^K\lambda(\sigma_{ik}) \|D_\theta(x_i{+}n_{ik}, \sigma_{ik}){-}x_i \|_2^2$
        \EndFor
        \State $G_{batch} = \frac{1}{B} \sum_{i \in \mathcal{B}} \texttt{clip}_C\left(\nabla_\theta \tilde{l}_i \right)$
        \State $\tilde{G}_{batch} = G_{batch} + (C/B) z, z \sim \mathcal{N}(\mathbf{0}, \sigma_\mathrm{DP}^2)$
        \State $\theta = \theta - \eta * \textit{Adam}(\tilde{G}_{batch})$
        \EndFor
    \end{algorithmic}
\end{algorithm}
\section{Additional qualitative results}
\label{app:qual-tobacco3482}
\begin{figure}[H]
    \centering
    \subfloat[$\varepsilon=1$]{%
        \includegraphics[width=\linewidth]{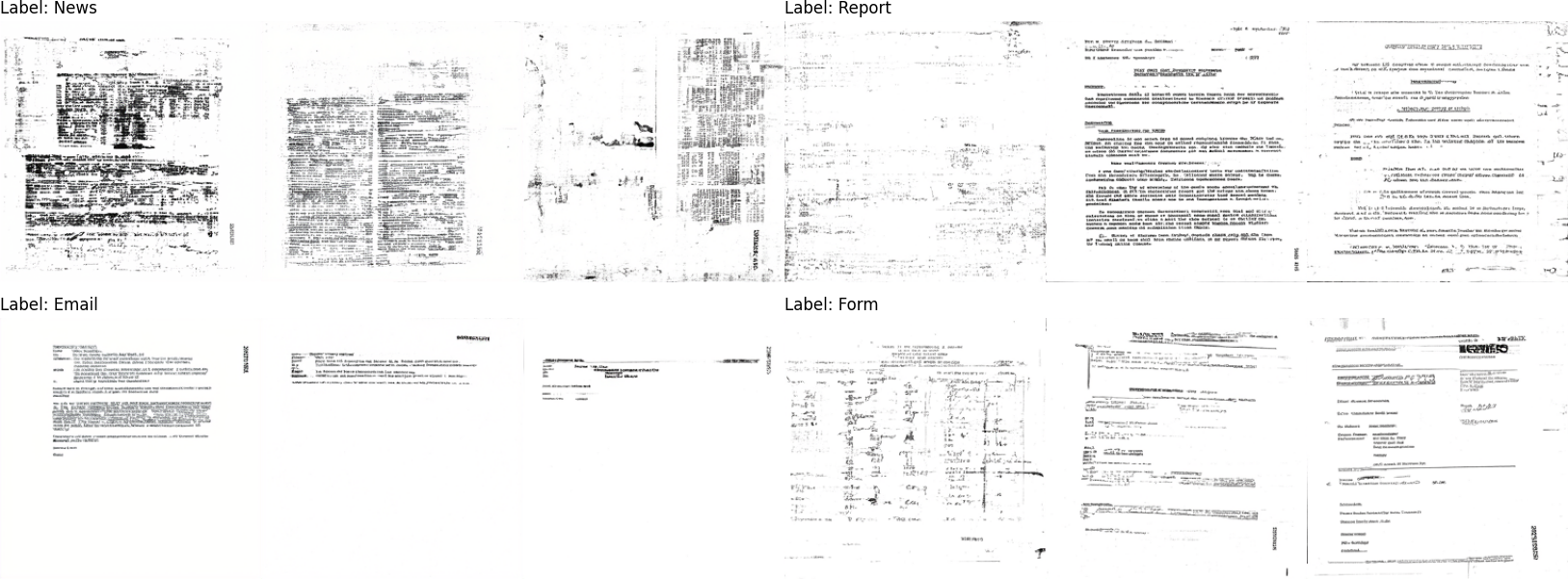}
        \label{fig:tobacco3482-eps-1}
    }
    \hfill
    \subfloat[$\varepsilon=5$]{%
        \includegraphics[width=\linewidth]{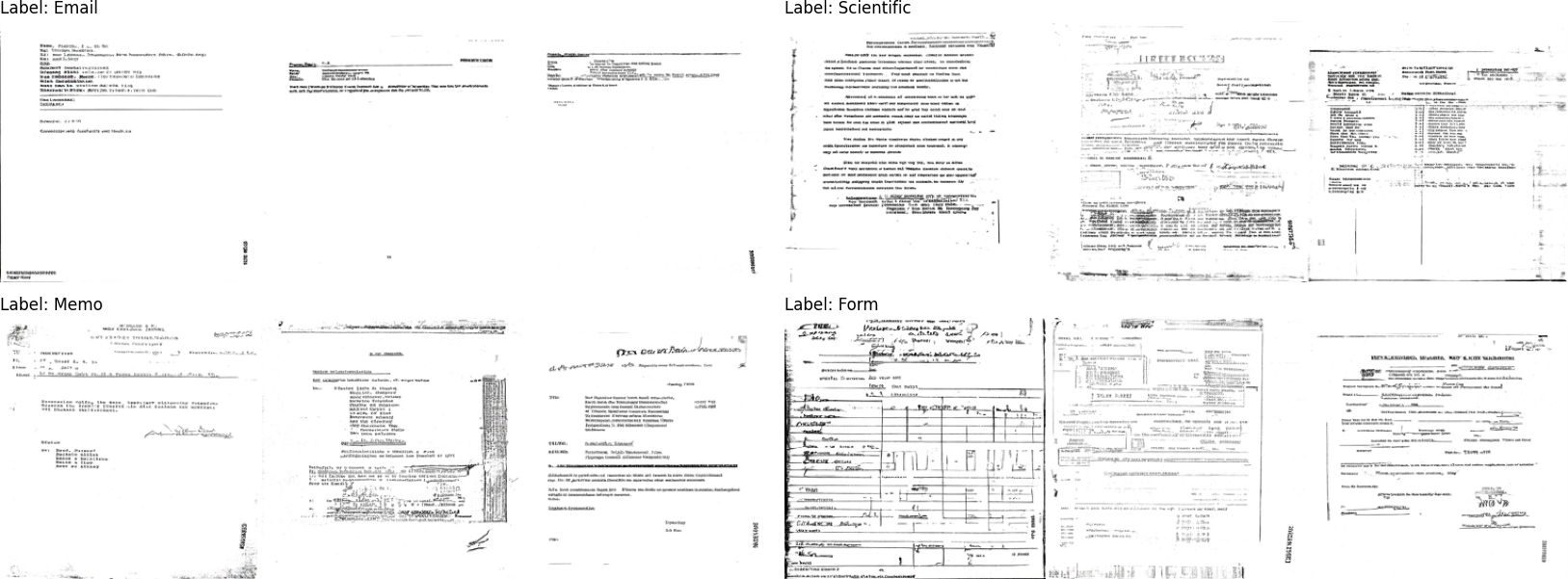}
        \label{fig:tobacco3482-eps-2}
    }
    \hfill
    \subfloat[$\varepsilon=10$]{%
        \includegraphics[width=\linewidth]{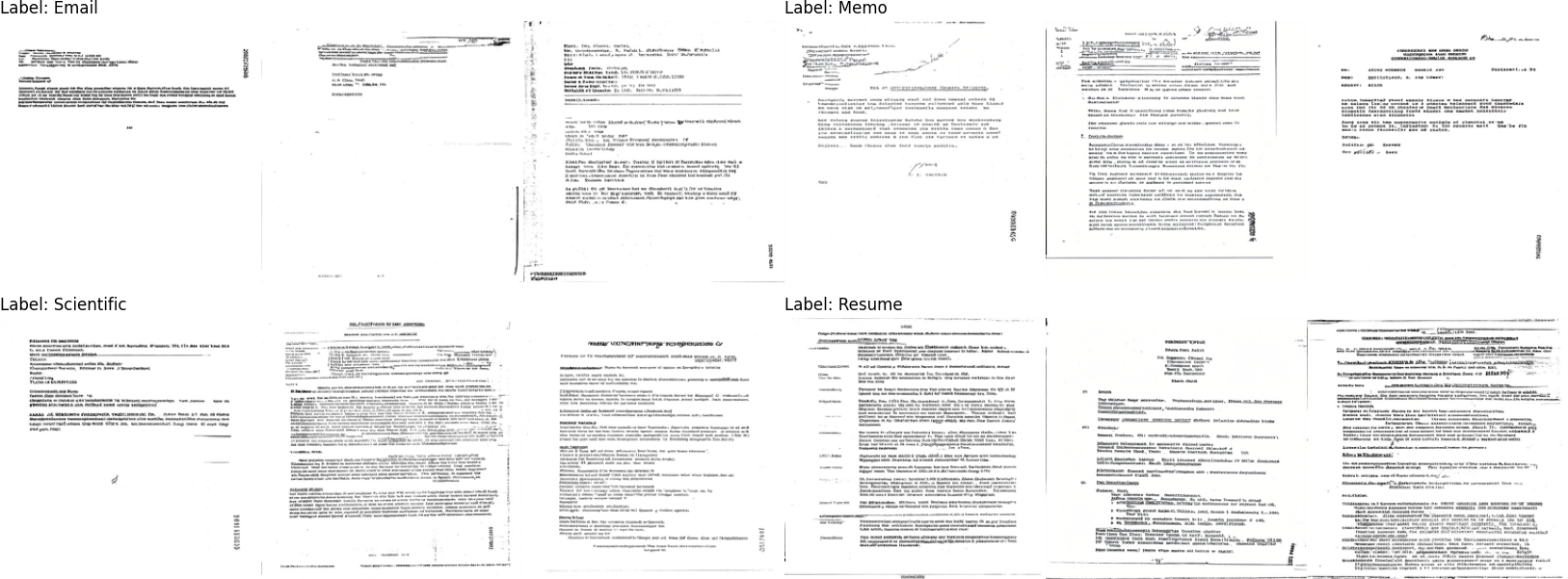}
        \label{fig:tobacco3482-eps-3}
    }
    \caption{
        Visual comparison of the synthetic Tobacco3482 dataset samples generated using KL-F4 autoencoder at $\varepsilon \in \{1, 5, 10\}$.
    }
    \label{fig:qual-tobacco3482}
    \vspace{-2em}
\end{figure}

\begin{figure}[H]
    \centering
    \subfloat[KL-F8]{%
            \includegraphics[width=\linewidth]{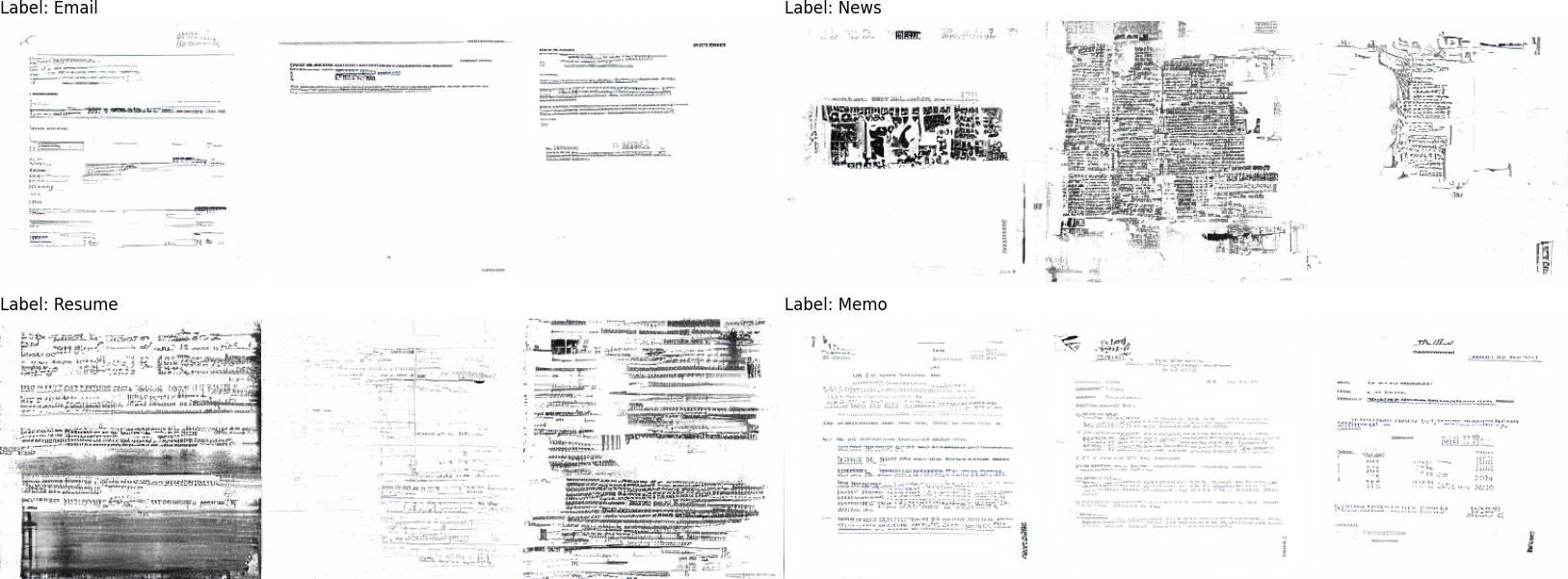}
            \label{fig:tob-eps-1-f4-vs-f8-1}
        }
    \hfill
    \subfloat[KL-F4]{%
            \includegraphics[width=\linewidth]{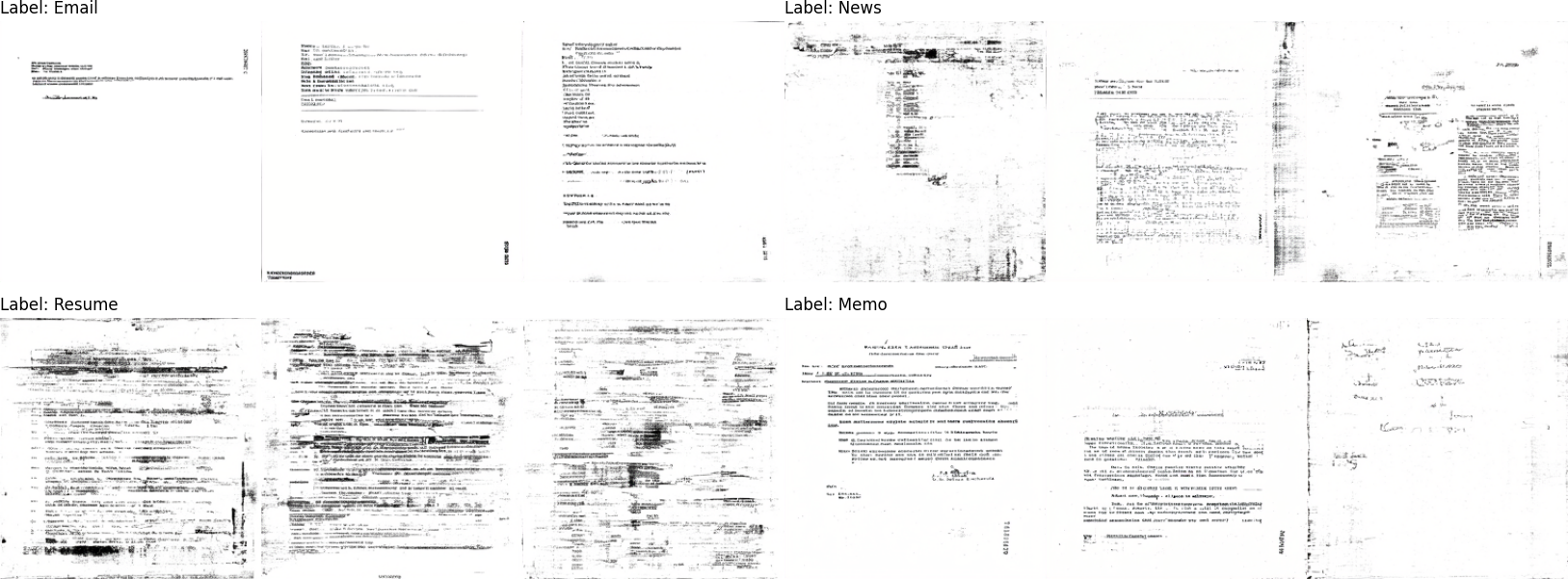}
            \label{fig:tob-eps--2f4-vs-f8-1}
        }
    \caption{
            Visual comparison of synthetic Tobacco3482 samples generated using KL-F8 and KL-F4 autoencoders at $\varepsilon = 1$.
            }
    \label{fig:qual-tobacco3482-comparison}
\end{figure}

In Fig.~\ref{fig:qual-tobacco3482}, we present qualitative results on the Tobacco3482 dataset at different privacy levels, $\varepsilon \in {1, 5, 10}$.
In Fig.~\ref{fig:qual-tobacco3482-comparison}, we present a qualitative comparison of the samples generated using the KL-F4 and KL-F8 autoencoder configurations on the Tobacco3482 dataset.
\end{document}